\documentclass{aa} 
\usepackage{natbib}

\usepackage{graphicx}
\usepackage{array}
\graphicspath{ {Figs/} }
\usepackage{multicol}
\usepackage{lipsum}
\usepackage{kantlipsum}
\usepackage{txfonts}
\usepackage{setspace}
\usepackage{amsmath}
\usepackage{hyperref}
\usepackage{lineno}
\usepackage{float} 
\begin{document} 

 \title{The quest for stable circumbinary companions to post-common envelope sdB eclipsing binaries}
   \subtitle{Does the observational evidence support their existence?}
   \author{D. Pulley\textsuperscript{1} 
          \and
          G. Faillace \textsuperscript{1} \
           \and
          D. Smith \textsuperscript{1}
           \and
          A. Watkins \textsuperscript{1}
           \and
          S. von Harrach \textsuperscript{1}
          }

   \institute{\textsuperscript{1}British Astronomical Association, Burlington House, Piccadilly, London W1J 0DU, UK\\
              \email{binarygfaillace@aol.com}\\
              \email{david@davidpulley.co.uk}
         }

   \date{Received: May 08, 2017 / Accepted: November 03, 2017 }

 
  \abstract
   {Period variations have been detected in a number of eclipsing close compact binary subdwarf B stars (sdBs) and these have often been interpreted as caused by circumbinary massive planets or brown dwarfs. According to canonical binary models, the majority of sdB systems are produced from low mass stars with degenerate cores where helium is ignited in flashes. Various evolutionary scenarios have been proposed for these stars, but a definite mechanism remains to be established. Equally puzzling is the formation of these putative circumbinary objects which must have formed either from the remaining post common envelope circumbinary disk or survived its evolution.  }
 {In this paper we review the eclipse time variations (ETVs) exhibited by seven such systems (EC10246-2707, HS0705+6700, HS2231+2441, J08205+0008, NSVS 07826147, NSVS 14256825 and NY Vir) and explore if there is conclusive evidence that the ETVs observed over the last two decades can reliably predict the presence of one or more circumbinary bodies.}
   {We report 246 new observations of the seven sdB systems made between 2013 September and 2017 July using a worldwide network of telescopes. We combined our new data with previously published measurements to analyse the ETVs of these systems.}
   {Our data shows that period variations cannot be modelled simply on the basis of circumbinary objects. This implies that more complex processes may be taking place in these systems.  These difficulties are compounded by the secondary star not being spectroscopically visible. From eclipse time variations, it has historically been suggested that five of the seven binary systems reported herein had circumbinary objects. Based on our recent observations and analysis only three systems remain serious contenders. We find agreement with other observers that at least a decade of observations is required to establish reliable ephemeris. With longer observational baselines it is quite conceivable that the data will support the circumbinary object hypothesis of these binary systems. Also we generally agree with other observers that larger values of (O-C) residuals are found with secondary companions of spectral type M5/6 or earlier possibly as a result of an Applegate type mechanism}
   {}


  \maketitle
%

\section{Introduction}

Since 2007, from observations of ETVs, there have been many claims made for the detection of circumbinary objects around post common envelope subluminous eclipsing binary systems.  These sdB systems are members of the HW Vir family of short period binary systems that consist of a very hot sdB type star and a cool, low mass, main sequence star or brown dwarf companion. Their compact structure and the large temperature difference between the two components give rise to short and well defined primary eclipses allowing times of minima to be determined with high precision. The sdB component of these systems have canonical masses of $\sim$~0.47$M_{\odot}$ and consist of a helium burning core with a thin hydrogen envelope and are located at the left hand extremity of the horizontal branch in the H-R diagram.  \\ \\
Various evolutionary scenarios have been proposed for these stars, but a definitive mechanism remains to be established, in particular whether or not binary evolution, as outlined by \cite{Pac}, \cite{Web} and  \cite{Zoro_a}, is a requirement. These models suggest that when the more massive primary evolves, it fills its Roche lobe during one of the red giant phases and unstable mass transfer from the primary occurs at a rate that cannot be accommodated by the secondary component.  This results in material forming a non co-rotating common envelope that surrounds the core of the red giant and the secondary component, thus enshrouding the binary system.  Angular momentum is transferred from the binary system to the common envelope bringing the binary pair closer together and resulting in a short binary period of typically between 2 and 3 hours.  Eventually the common envelope has sufficient orbital energy to overcome its binding energy and is mostly ejected from the system in a timescale of $\sim$1000 years \citep{Xiong} initially creating a proto-planetary nebula then a planetary nebula leaving a stellar remnant well on its way to a white dwarf.  The short duration of the common envelope phase means the mass of the secondary companion is assumed to remain constant and the mass of the remaining compact object, the primary, would be effectively equal to the mass of the core of the red giant at the onset of mass transfer.\\\\
\cite{Zoro_c} provided an overview of thirteen of these systems. Interestingly, in five of the thirteen systems’ eclipse ETVs have been interpreted as showing the presence of low mass circumbinary objects e.g. brown dwarfs, massive planets etc.  If such bodies do exist, then they must have either survived the energetic common envelope ejection process, or have formed during this short period from the remnant of the ejected common envelope.  From simulations, \cite{Zoro_c},  concluded that the latter was most likely, the so-called second generation hypothesis.  Supporting this view, \cite{Schlei} concluded that the formation of planets from the ejected common envelope seemed feasible. Using primarily the NN Ser post common envelope eclipsing binary system model, which has confirmed circumbinary planets \citep{Pars,Hardy}, they estimated the mass loss and the fraction of mass that remains gravitationally bound to the system. This remaining mass would form a proto-planetary disk, resulting in new planets. In contrast, \cite{bear2014first}, based on angular momentum considerations, proposed a first generation hypothesis, in which existing planets survive the energetic common envelope ejection process.\\\\
In this paper we present the eclipse time variations exhibited by the seven sdB systems and analyse the results in the context of the circumbinary planet hypothesis. The seven systems studied are listed in Tables 1 and 2.
\section{Observing method and data reduction}
 We report 246 new observations of the seven sdB systems made between 2013 September and 2017 July using the telescopes and filters listed in Table A.1 and A.2\footnote{Tables A.1–A.3 are available in electronic form at the CDS via anonymous ftp to cdsarc.u-strasbg.fr (130.79.128.5) or via http://cdsweb.u-strasbg.fr/cgi-bin/qcat?J/A+A/}.  The effects of differing atmospheric extinctions were minimised by making all observations at altitudes of generally greater than 40$^{\circ}$.  All images were calibrated using dark, flat and bias frames and then analysed with MaxIm DL software\footnote{MaxIm DL,\url{http://www.diffractionlimited.com/}}. The source flux was determined with aperture photometry using a variable aperture, whereby the radius is scaled according to the full width at half-maximum (FWHM). Variations in observing conditions were accounted for by determining the flux relative to a comparison star in the field of view. Apparent magnitudes, coordinates and separation from the target for each comparison star are given in Table A.3. \\\\
 When using filters the targets’ apparent magnitude was derived from the apparent magnitudes of the comparison stars and the average magnitude of the target calculated by the software. The comparison stars catalogue magnitudes for various filters were taken from the American Association of Variable Star Observers (AAVSO) Photometric All Sky Survey (APASS) catalogue of similar magnitudes as the targets.  An attempt was made to select comparison stars with similar colour indices as the target stars, although this proved difficult given the very blue nature of the sdB systems. Because the APASS catalogue does not include the R pass band, in the few cases where observations were taken with the R filter a conversion formula recommended by AAVSO was used to transform the catalogue Sloan r’ magnitudes to corresponding R magnitudes.When observations were performed without filter,  check stars were used to ensure that there was no variability in the reference star selected. A summary of the various target systems investigated together with their basic properties are listed in Tables 1 and 2.  A detailed analysis of each system is provided in the following sections. \\
\noindent\\
All of our new timings used in this analysis were first converted to Barycentric Julian Date Dynamical Time $(BJD\_TBD)$ using the time utilities of the Ohio State University\footnote{Ohio State University, \url{http://astroutils.astronomy.ohio-state.edu/time/}}. We then calculated the times of minima using the  \cite{Kwee} procedure coded in the Peranso \footnote{Peranso, \url{http://www.peranso.com/}}  software package and cross checked these results with both Kwee and van Woerden and the Fourier procedures of the Minima\footnote{Nelson B., Minima,  \url{https://www.variablestarssouth.org/software-by-bob-nelson/}}  software package.  In one instance, EC 10246-2707 at JD 2457408.6459, we had only observations around the minimum and through the egress. Here we fitted an inverted Gaussian curve, with parameters derived from a good light curve, minimising the residuals and deriving a time of minimum which was in very good agreement with the expected time of minimum.\\
 \begin{table*}[ht]
 \renewcommand\thetable{1}
\caption{Summary of the seven objects observed.  A total of 246 times of minima have been determined of which 221 were primary minima.
}             
\label{table:4} 
\centering
{\footnotesize
\begin{tabular}
{p{0.13\linewidth}p{0.10\linewidth}p{0.11\linewidth}p{0.13\linewidth}p{0.03\linewidth}p{0.06\linewidth}p{0.07\linewidth}p{0.03\linewidth}p{0.2\linewidth}p{0.3\linewidth}}
\hline
\\
\multicolumn{1}{c}{Object} & \multicolumn{1}{c}{RA} & \multicolumn{1}{c}{Dec} & \multicolumn{1}{c}{Period} &\multicolumn{1}{l}{Distance} & \multicolumn{1}{c}{Mag} & \multicolumn{1}{c}{Pri} & \multicolumn{1}{c}{Sec} & \multicolumn{1}{c}{Observing}\\
\multicolumn{1}{c}{} & \multicolumn{1}{c}{J2000} & \multicolumn{1}{c}{J2000} & \multicolumn{1}{c}{(days)} & \multicolumn{1}{c}{kpc} &&  \multicolumn{1}{c}{Min} & \multicolumn{1}{c}{Min} & \multicolumn{1}{c}{Period}
\\\\
\hline
\\EC 10246-2707 & 10 26 56.472 & -27 22 57.11 &\multicolumn{1}{l}{0.118507985} &\multicolumn{1}{l}{-} & 14.77(V) & \multicolumn{1}{r}{7} & \multicolumn{1}{r}{0} & \multicolumn{1}{c}{2016 Jan - 2017 Jan} \\
HS 0705+6700 & 07 10 42.056  & +66 55 43.52 & \multicolumn{1}{l}{0.095646609}& \multicolumn{1}{l}{1.7} & 14.60(R) & \multicolumn{1}{r}{101} & \multicolumn{1}{r}{16} & \multicolumn{1}{c}{2013 Sep- 2017 May} \\ 
	HS 2231+2441 & 22 34 21.483 & +24 56 57.39 &\multicolumn{1}{l}{0.110587829}&\multicolumn{1}{l}{0.17} & 14.20(V) & \multicolumn{1}{r}{26} & \multicolumn{1}{r}{3} & \multicolumn{1}{c}{2014 Dec- 2017 Jan} \\ 
	J08205+0008 & 08 20 53.076 & +00 08 42.78 &\multicolumn{1}{l}{0.096240737} &\multicolumn{1}{l}{0.61} & 15.17(V) & \multicolumn{1}{r}{33} & \multicolumn{1}{r}{6} & \multicolumn{1}{c}{2014 Dec - 2017 Feb}  \\ 
	NSVS 07826147 & 15 33 49.446 & +37 59 28.06 & \multicolumn{1}{l}{0.1617704491} & \multicolumn{1}{l}{0.6} & 13.08(V) & \multicolumn{1}{r}{20} &\multicolumn{1}{r}{0} & \multicolumn{1}{c}{2015 Apr - 2017 Jun}  \\ 
	NSVS 14256825 & 20 20 00.475 & +04 37 56.49 & \multicolumn{1}{l}{0.1103741681} & \multicolumn{1}{l}{0.6} & 13.34(R) & \multicolumn{1}{r}{19} & \multicolumn{1}{r}{0} & \multicolumn{1}{c}{2015 Aug- 2017 Jul }\\ 
	NY Vir & 13 38 48.142 & -02 01 49.24 & \multicolumn{1}{l}{0.1010159668} & \multicolumn{1}{l}{1.1}  & 13.30(V) & \multicolumn{1}{r}{15} & \multicolumn{1}{r}{0} & \multicolumn{1}{c}{2015 Apr - 2017 Jun} \\
\hline
\end{tabular}}
\end{table*}
\\
  \begin{table*}[ht]
\renewcommand\thetable{2}
\caption{Summary of key parameters of the binary systems observed.  The spectral types for theses systems are not clearly defined being indirectly determined from light curve paramenters which themselves can be poorly constrained.  The RMS of residuals for NY Vir is shown with two values.  Utilising all data the value is 36.45 but eliminating the two Vuckovic data points which Lee et al. (2014) reports with very low uncertainties yields a value of 4.78.}             
\label{table:5} 
\centering
\resizebox{\textwidth}{!}
{\footnotesize
\begin{tabular}             
{p{0.12\linewidth}p{0.04\linewidth}p{0.04\linewidth}p{0.03\linewidth}p{0.04\linewidth}p{0.05\linewidth}p{0.04\linewidth}p{0.03\linewidth}p{0.05\linewidth}p{0.03\linewidth}p{0.05\linewidth}p{0.17\linewidth}}
\hline\\
Object & \multicolumn{1}{c}{$M_{1}$}  & \multicolumn{1}{c}{$M_{2}$} & \multicolumn{1}{c}{incl} & \multicolumn{1}{c}{$T_{eff}$ } & \multicolumn{1}{c}{a} & \multicolumn{1}{c}{Sp.Type}& \multicolumn{1}{c}{RMS}& logg& \multicolumn{1}{c}{K1} & $log(\frac{n_{He}}{n_H})$ & \multicolumn{1}{c}{Reference}\\
&\multicolumn{1}{c}{($M_{\odot}$)}&\multicolumn{1}{c}{($M_{\odot}$)}&\multicolumn{1}{c}{(degs)}&\multicolumn{1}{c}{(K)}&\multicolumn{1}{c}{($R_{\odot}$)}&\multicolumn{1}{c}{($M_{2}$)}&\multicolumn{1}{c}{(Residuals)}&&\multicolumn{1}{c}{(km/s)}&&\\\\
\hline
 \\EC 10246-2707  & 0.45 & 0.12& \multicolumn{1}{c}{80.0} & 28900&0.84&\multicolumn{1}{c}{M5/M6}&\multicolumn{1}{c}{1.60} & 5.64&71.6&-2.50&Barlow et al. (2013)\\
  HS 0705+6700  & 0.48 & 0.13 & \multicolumn{1}{c}{84.4} & 28800 & 0.81 &\multicolumn{1}{c}{M4/M5}&\multicolumn{1}{c}{19.89} & 5.40 & 85.8 & -2.68 & Drechsel et al. (2001) \\ 
HS 2231+2441  & 0.265 & 0.05 & \multicolumn{1}{c}{79.1} & 28370 & 1.18&\multicolumn{1}{c}{BD}&\multicolumn{1}{c}{2.47}  & 5.39 & 49.1 & -2.91 & Ostensen et al. (2008) \\ 
	J08205+0008 & 0.25 & 0.045 & \multicolumn{1}{c}{85.9} & 26700 & 0.6 &\multicolumn{1}{c}{BD}&\multicolumn{1}{c}{1.13} & 5.48 & 47.4 &  & Geier et al. (2011) \\ 
    &/0.47&/0.068&&&/1.1&&&&\\
	NSVS 07826147& 0.376 & 0.113 & \multicolumn{1}{c}{86.6} & 29230 &  &\multicolumn{1}{c}{M5}&\multicolumn{1}{c}{2.41} & 5.58 & 71.1 & &  For et al. (2010) \\ 
	NSVS 14256825 & 0.35 & 0.097 & \multicolumn{1}{c}{82.5} & 42300 & 0.74&\multicolumn{1}{c}{M5/M7}&\multicolumn{1}{c}{20.95}  & 5.50 & 73.4 & -2.52 & Almeida et al. (2012)\\
    &/0.46&/0.21&&/35250&&&&/5.49&&/-2.70&Kilkenny \& Koen (2012) \\ 
	NY Vir & 0.39 & 0.11 & \multicolumn{1}{c}{80.7} & 31300 & 0.72 &\multicolumn{1}{c}{M5} &\multicolumn{1}{c}{4.78/36.45} &  5.74 & 78.6 & -2.91 & Vuckovic et al. (2007) \\
\hline
\end{tabular}}
\end{table*}
\\Our new timings were combined with previously published times of minima and, where appropriate, the historic times were converted to $BJD\_TBD$ before computing a new ephemeris and/or residuals.  Where a new linear or quadratic ephemeris was calculated only observed primary minima data was used.  We also excluded data that used phase folded or synthetic light curves techniques.  These techniques are frequently employed in data analysis of wide field sky surveys.  \\\\
\noindent
The difference between the observed and calculated times of minima (O - C) can be used to infer potential internal or external influences on the binary pair, for example (i) angular momentum loss through magnetic breaking or the emission of gravitational waves (ii) angular momentum redistribution through an Applegate type mechanisms ,(iii) the apparent changing of the binary period through the presence of a circumbinary object or (iv) apsidal motion. See for example \cite{brink}, \cite{bours2016long} and references therein\\
\section{Analysis of Eclipse timings}
The (O – C) diagrams for each system shown in this section include both primary and secondary minima together with data from wide field sky surveys such as the Northern Sky Variability Survey (NSVS).  SuperWASP data has been included where this fills significant gaps in the historic data; when included we used the same binning methodology as \cite{Lohr}.  Error bars are shown where timing uncertainties have been included in published data and did not affect the readability of the diagram.\\\\
\noindent
In the following subsections, we present both a brief historical review  of seven sdB eclipsing binary systems, reviewing the hypotheses presented by earlier observers followed by an analysis of our new results.  All our new times of minima for these systems are listed in Table A.2. 

 \subsection{EC10246-2707}

\subsubsection{Background:} EC10246-2707 was identified as an sdB star as part of the Edinburgh-Cape Faint Blue Object Survey \citep{Kilk_a} and discovered, during follow-up observations, to be an eclipsing binary system with a period of $\sim$2.8hr.  Simultaneously, and independently, Barlow  identified the binary nature of the star. A A joint University of North Carolina/Edinburgh-Cape paper was published, \cite{Barl}, reporting 49 times of minima and embracing spectroscopic and light curve analysis together with a linear ephemeris covering 15 years of observations.  They determined the period of the binary as 0.1185079936 days and identified this system as a typical member of HW Vir family with T$_{\text{1eff}}$= 28,900K, logg = 5.64, $M_{1}$ = 0.45$M_{\odot}$  and a likely M dwarf companion of mass 0.12$M_{\odot}$.  They explored the possibility of a varying binary period but found no statistical evidence to support this as opposed to a constant period. \\
\noindent\\
\cite{Kilk_e} provided ten new primary minima timings between 2013 and 2014 using the 1m telescope at the South African Astronomical Observatory.  He calculated a new linear ephemeris:
\begin{equation}
\hspace{0.25cm} T_{min,BJD} = 2450493.46727(3) + 0.1185079951(7) * E
\end{equation}

As noted by \cite{Kilk_e} the small change in residuals around mid 2012 lends support to a time varying period:

\begin{equation}
\begin{split}
\hspace{0.25cm} T_{min,BJD}=2450493.46733(3) + 0.118507985(3) *E\\ + 2.0(5) * 10^{-13} * E^2
\end{split}
\end{equation}
\noindent
\cite{Kilk_e} also showed that the effect of gravitational radiation on binary period reduction in HW Vir like systems is only detectable, with present equipment accuracies, after approximately a century and so could not explain these observed short period changes.\\
Application of the Mann-Whitney U test (to identify whether the ranked absolute values of residuals from both the linear and quadratic ephemerides could come from the same parent population) indicates, at the p = 0.05 level, that the linear ephemeris does not provide a statistically better fit to the data than the quadratic ephemeris. More data providing a longer timebase will verify which ephemeris is most applicable. 
\subsubsection{Recent data and ephemeris:}We have made seven observations between 2016 January and 2017 January using the 0.43m iTelescopes T17 Siding Spring, the 1m Cerro Tololo and the 1m Sutherland telescopes.  Our results integrate well with earlier data from \cite{Kilk_e} and confirm an apparent change in binary period.
\begin{figure}[htbp]
\centering
{\includegraphics[width=\linewidth]{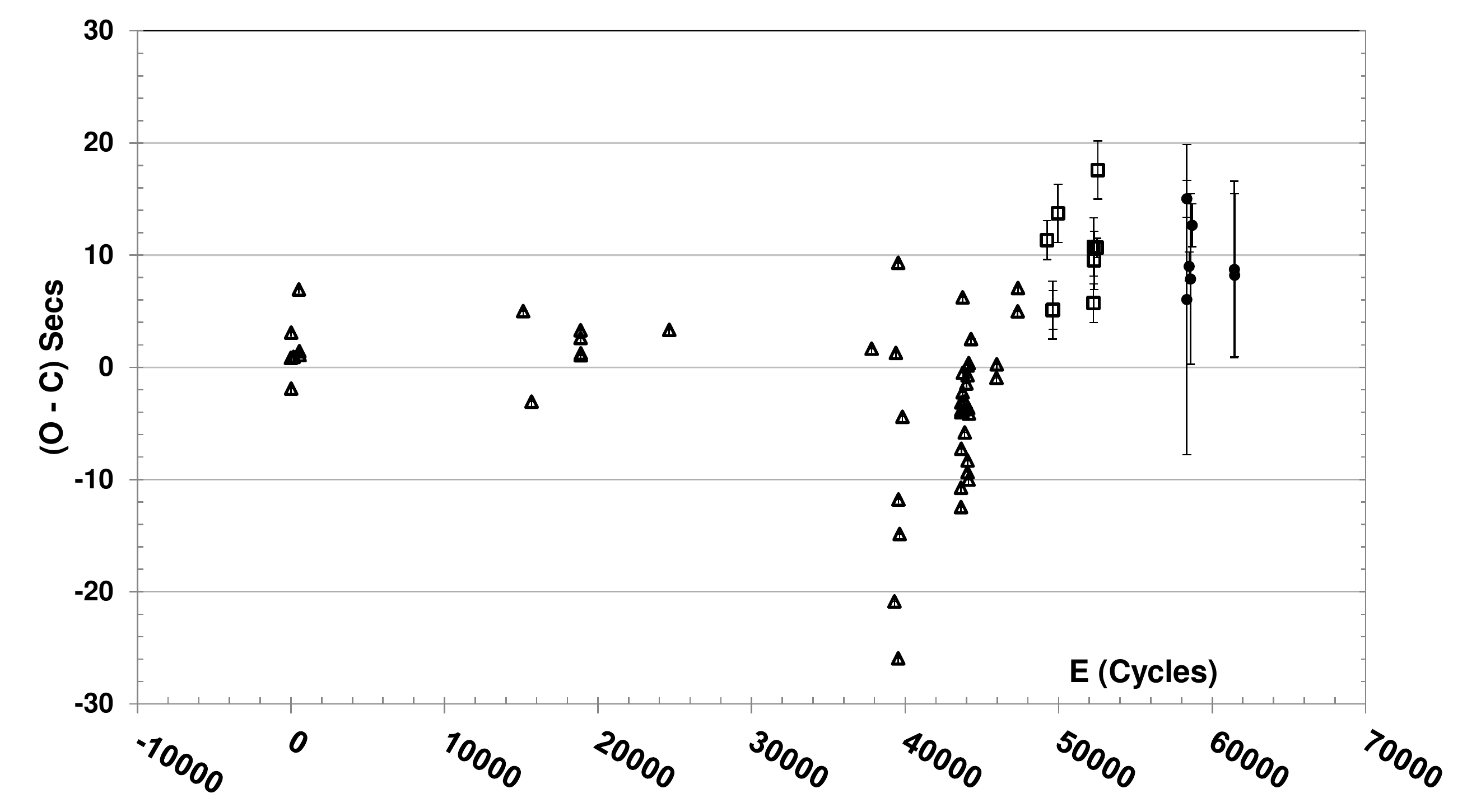}}
\caption{(O – C) diagram for EC10246-2707 using Kilkenny’s (2014) linear ephemeris with our new data shown as  solid dots; \cite{Kilk_e} data is shown as open squares; open triangles are \cite{Barl} data}
\label{fig1}
\end{figure}
\noindent\\\\
With our new data, together with published historical data, we re-calculated both the linear and quadratic ephemerides, adopting the same reference epoch as used in \cite{Kilk_e}.  Since we found little difference between this ephemeris and our derived ephemeris, we continue to use Kilkenny’s ephemeris.  The (O – C) diagrams with our new data are shown in Figs. 1 and 2 for the linear and quadratic ephemerides respectively.  The linear fit of Fig. 1 does indicate a small binary period increase over the 20 year observational time span but this increase appears to commence at E $\sim$47,000, corresponding to mid 2012. \\ \\
To quantify the goodness of fit we determine the RMS of the weighted residuals (sometimes referred to as the reduced $\chi^2$) defined as:
\begin{equation}
\begin{split}
\hspace{2cm} RMS =\sqrt{\frac{1}{n} \sum_{i}^{n} \bigg(\frac{O_i-C_i}{\sigma_i}\bigg)^2  }
\end{split}
\end{equation}

\noindent
in which $O_i$, $C_i$ and $\sigma_1$ are the observed, calculated and uncertainty in eclipse time at cycle i taken over the n data sets. For the linear and quadratic ephemerides the RMS computes to 3.02 and 1.60 respectively.\\\\
Also of note is the wide spread in residuals, approximately 35s, at E $\sim$39,500 and which is very much greater than the typical error bars associated with each minimum.  To a lesser degree this can also be seen at other times e.g. E $\sim$ 44,000.  More data will be required to establish the underlying trends.
\begin{figure}[htbp]
\centering
{\includegraphics[width=\linewidth]{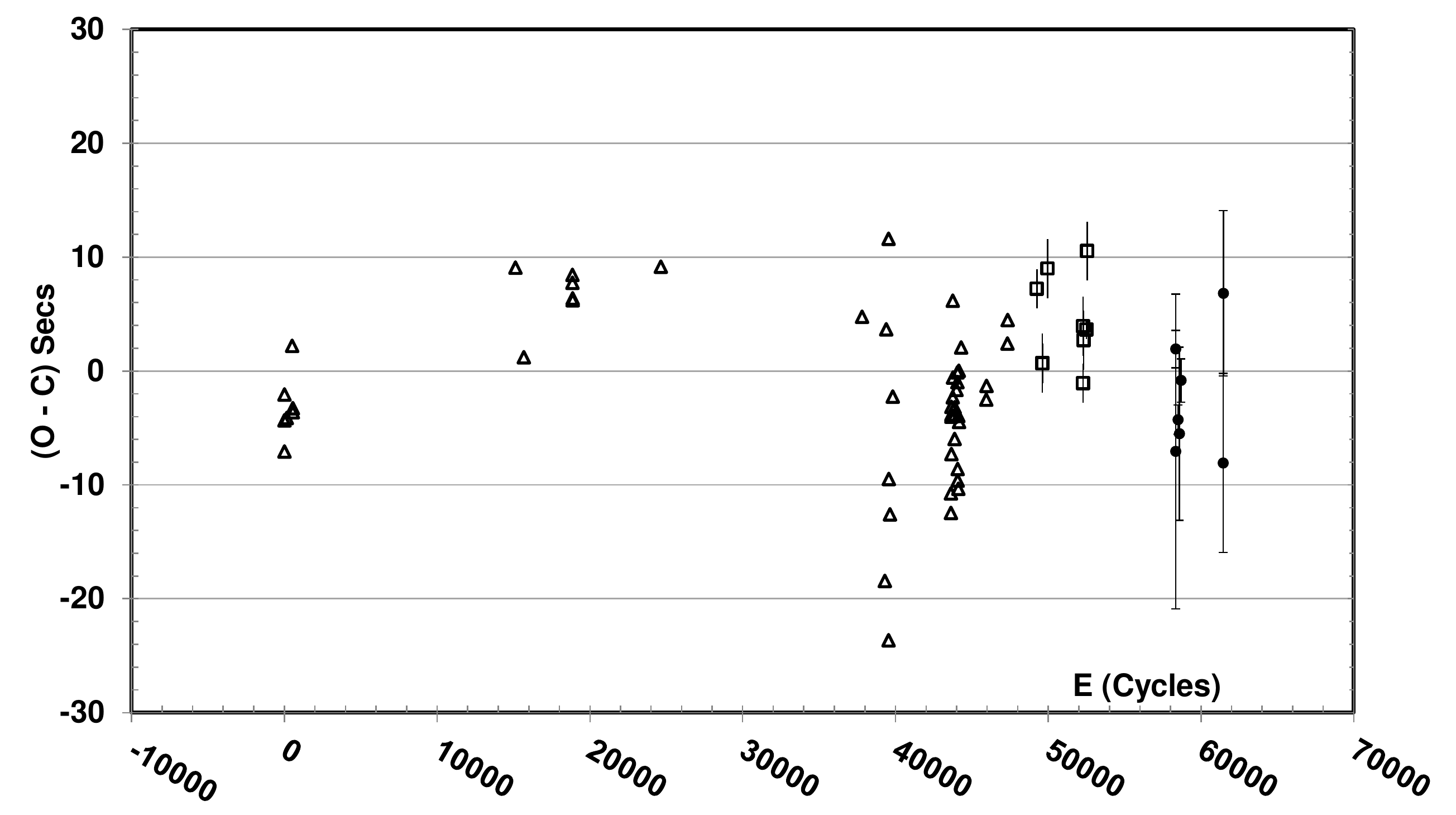}}
\caption{(O – C) diagram for EC10246-2707 using Kilkenny’s (2014) quadratic ephemeris with our new data shown as solid dots; open squares is Kilkenny’s data; open triangles are Barlow and Kilkenny's 2013 data.}
\label{fig2}
\end{figure}\\
\subsection{HS0705+6700}
\subsubsection{Background:}
 HS0705+6700 was identified as a 14.7 magnitude sdB star from the Hamburg Schmidt Quasar Survey.  Follow-up observations by \cite{Drech} showed this to be a short period, $\sim$2.3hr, post common envelope eclipsing binary system and they produced the first linear ephemeris for this system.  From their light curve and radial velocity measurements they calculated the mass and radius of the primary and secondary components to be 0.483$M_{\odot}$ and 0.239$R_{\odot}$ and  0.134$M_{\odot}$ and 0.186$R_{\odot}$ respectively.  The orbital semi-major axis and inclination were determined as 0.81$R_{\odot}$ and $84.4^\circ$ respectively.  The effective temperature of the sdB component was estimated to be 28,800 $\pm$900K.  From these values they concluded the close companion was a cool M dwarf star contributing significant amounts of phase dependant reflected light. \\ 
  
\noindent 
Whilst \cite{Nia} and \cite{Cam} provided confirmatory light curve parameters, most investigators \citep{Qian_a,Qian_b,Qian_d,Cam,Beu,Pull}  have focused on an apparent quasi-periodic variation in the eclipsing binary period.  Although not proven, and following the rejection of other possible mechanisms, e.g. orbital period modulation \citep{Appg} or apsidal motion, the most likely cause of this variation has been attributed to the presence of a circumbinary object.  A summary of the findings of these investigators is presented in Table 3.\\
 \\
  \begin{table*}[ht]
\renewcommand\thetable{3}
\caption{A summary of observations and third body orbital parameters reported by previous observers for HS0705+6700.  Note that the circumbinary period reported by Qian et al. (2010) should probably read 7.15 years. We think this is a typographical error.}             
\label{table:6} 
\centering
{\footnotesize
\begin{tabular} 
{p{0.19\linewidth}p{0.02\linewidth}p{0.08\linewidth}p{0.04\linewidth}p{0.1\linewidth}p{0.08\linewidth}p{0.09\linewidth}p{0.09\linewidth}p{0.1\linewidth}}
\hline
& \multicolumn{1}{c}{Drechsel} & \multicolumn{1}{c}{Niarchos}& \multicolumn{1}{c}{Qian}&\multicolumn{1}{c}{Qian}&\multicolumn{1}{c}{Carmurdan}&\multicolumn{1}{c}{Beuermann}&\multicolumn{1}{c}{Qian}&\multicolumn{1}{c}{Pulley} \\   
\multicolumn{1}{c}{Parameter} &\multicolumn{1}{c}{et al.}&\multicolumn{1}{c}{et al.}&\multicolumn{1}{c}{et al.}&\multicolumn{1}{c}{et al.}&\multicolumn{1}{c}{et al.}&\multicolumn{1}{c}{et al.}&\multicolumn{1}{c}{et al.}&\multicolumn{1}{c}{et al.}\\
&\multicolumn{1}{c}{(2001)}&\multicolumn{1}{c}{(2003)}&\multicolumn{1}{c}{(2009)}&\multicolumn{1}{c}{(2010)}&\multicolumn{1}{c}{(2012)}&\multicolumn{1}{c}{(2012)}&\multicolumn{1}{c}{(2013)}&\multicolumn{1}{c}{(2015)}\\
\hline  
Radial velocity analysis & \multicolumn{1}{c}{Yes} & \multicolumn{1}{c}{No} & \multicolumn{1}{c}{No} & \multicolumn{1}{c}{No} & \multicolumn{1}{c}{No} & \multicolumn{1}{c}{No} &\multicolumn{1}{c}{No} & \multicolumn{1}{c}{No} \\ 
Light curve analysis &\multicolumn{1}{c}{Yes} & \multicolumn{1}{c}{Yes} & \multicolumn{1}{c}{No} & \multicolumn{1}{c}{No} &\multicolumn{1}{c}{Yes} & \multicolumn{1}{c}{No}& \multicolumn{1}{c}{No} & \multicolumn{1}{c}{No}  \\ 
Primary eclipse & \multicolumn{1}{c}{13} & \multicolumn{1}{c}{3} & \multicolumn{1}{c}{33} & \multicolumn{1}{c}{12} & \multicolumn{1}{c}{6} & \multicolumn{1}{c}{19} & \multicolumn{1}{c}{73} & \multicolumn{1}{c}{40}\\
      Secondary eclipse & \multicolumn{1}{c}{0} & \multicolumn{1}{c}{3} & \multicolumn{1}{c}{5} &\multicolumn{1}{c}{2} &\multicolumn{1}{c}{0} & \multicolumn{1}{c}{3} & \multicolumn{1}{c}{5} & \multicolumn{1}{c}{12} \\
Binary ephemeris &  &  &  &  &  &  &  &  \\\hline
Binary period (days) &\multicolumn{1}{r}{\tiny 0.09564665} & as Dreschel & \tiny 0.095646625  & \multicolumn{1}{r}{\tiny0.095646684}   & \tiny 0.095564665 & \tiny 0.0956466253 & \tiny 0.095946611 &  \multicolumn{1}{r}{\tiny0.0956466012}   \\ 
Quadratic term (days) & \multicolumn{1}{c}{-} & \multicolumn{1}{c}{-} & \multicolumn{1}{c}{-} & \multicolumn{1}{c}{$-1.87 x 10^{-12}$} & \multicolumn{1}{c}{-} & \multicolumn{1}{c}{-} & \multicolumn{1}{c}{$2.57 x 10^{-12}$}&\multicolumn{1}{c}{$1.52 x 10^{-12} $}\\ 
Circumbinary Period (Yrs) & \multicolumn{1}{c}{-} & \multicolumn{1}{c}{-} & \multicolumn{1}{c}{7.15} & \multicolumn{1}{c}{15.7} & \multicolumn{1}{c}{8.06} & \multicolumn{1}{c}{8.41} & \multicolumn{1}{c}{8.87} & \multicolumn{1}{c}{8.73}\\ 
Eccentricity &\multicolumn{1}{c}{-} & \multicolumn{1}{c}{-} & \multicolumn{1}{c}{0} & \multicolumn{1}{c}{0} & \multicolumn{1}{c}{not given} & \multicolumn{1}{c}{0.38} & \multicolumn{1}{c}{0.19} & \multicolumn{1}{c}{0.22} \\
Semi-amplitude (s) & \multicolumn{1}{c}{-} & \multicolumn{1}{c}{-} & \multicolumn{1}{c}{92.4} & \multicolumn{1}{c}{71} & \multicolumn{1}{c}{98.5} & \multicolumn{1}{c}{86} & \multicolumn{1}{c}{87.4} & \multicolumn{1}{c}{80.1} \\
Min mass 3rd body$(M_J)$ & \multicolumn{1}{c}{-} & \multicolumn{1}{c}{-} & \multicolumn{1}{c}{39.5} & \multicolumn{1}{c}{30.4} & \multicolumn{1}{c}{37.7} & \multicolumn{1}{c}{29.6} & \multicolumn{1}{c}{32} & \multicolumn{1}{c}{33.2} \\
\hline
\end{tabular}}
\end{table*}
\normalsize
\cite{Qian_b} predicted a period of 15.7 years for a circumbinary object in a circular orbit around the binary system.  However, their Eq. 2 suggests that this is a typographical error and most probably the period should read 7.15 years.  They also found that the residuals could be further minimised with the introduction of a negative quadratic term into the ephemeris which they suggested could indicate the presence of a second circumbinary object. \cite{Cam} identified a third light in their light curve analysis which, they believed, supported the circumbinary object hypothesis.  However neither they, nor \cite{Beu}, could find evidence to support a long-term period decrease.\\
\noindent\\
Data from \cite{Qian_d} and \cite{Pull} extended the time line of the apparent circumbinary object into its second orbit and both preferred the inclusion of a positive quadratic term in the ephemeris.

\subsubsection{Recent data and ephemeris:}
In the Addendum to \cite{Pull} it was noted that a further 27 measurements taken between 2014 October and 2015 April showed a small departure from the predicted sinusoid of the third body.  Insufficient data had been gathered to confirm both the validity and significance of this result but we can now report a further 61 times of primary minima and four secondary minima observed between 2015 October and 2017 May.\\ 
\noindent\\
From an analysis of all known times of minima between 2000 February and 2014 April we published an unweighted quadratic ephemeris of:
\begin{equation}
\begin{split}
\hspace{0.25cm}T_{min,BJD}=2451822.76155(5) + 0.095646609(4)* E\\+ 5.5(9)*10^{-13}* E^2 + \tau{_3}
\end{split}
\end{equation}
The cyclical light travel time effect, $\tau_3$, is given by the equation of Irwin, see \cite{Irwin}:\\
\begin{equation}
\hspace{0.25cm} \tau_3 =\frac{a_{12}sin\hspace{2pt}i}{c}\Bigg[(1-e^2)\frac{sin(\nu+\omega)}{1+ecos\nu}+esin\omega\Bigg] 
\end{equation}
where the projected semi-major  axis, $a_{12}sini$, is 0.177; orbital eccentricity, e, is 0.03, longitude of periastron, $\omega$, is 0.119 and $\nu $ is the true anomaly. Full details can be found in Table 4,  \citet{Pull}. Using the quadratic portion of the above ephemeris we compute an RMS value of the weighted residuals of 19.89.\\
\noindent\\
We have computed the (O – C) residuals for our new data (E > 57000) and plotted these against the quadratic portion of the ephemeris of Eq. 4, as shown in Fig. 3.  The cyclical nature of the residuals computed from Eq. 5 is shown by the sinusoidal like curve in Fig. 3.  We can confirm that the observations post E = 55000 ($\sim$2015 February) depart from predicted values as first reported in the Addendum to our 2015 paper. This effect can be clearly seen in Fig. 4 where the residuals have been computed after removing both the quadratic and cyclical effects described by Eq. 4.\\
\begin{figure}[h]
\centering
{\includegraphics[width=\linewidth]{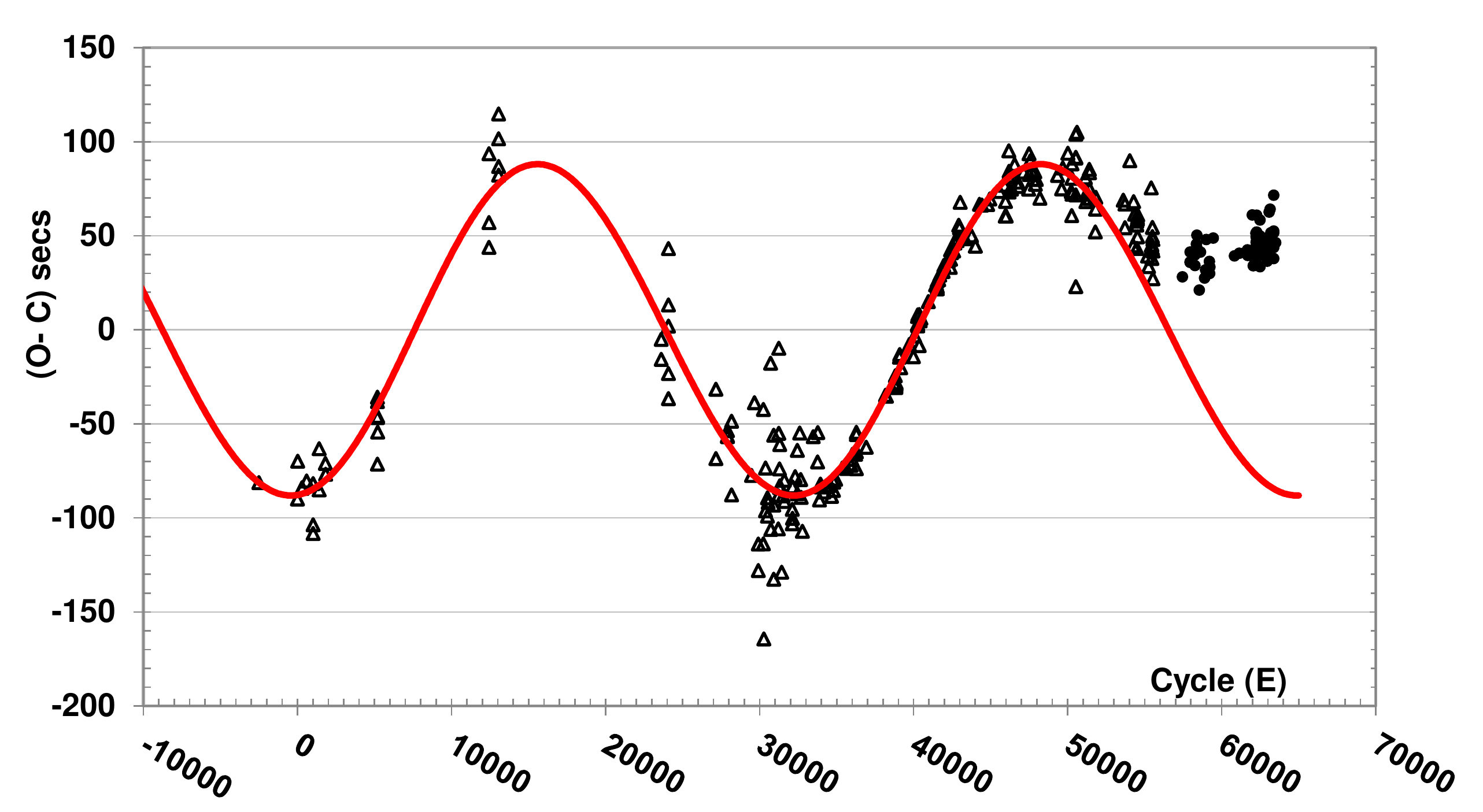}}
\caption{(O - C) residuals for HS0705+6700 after removing the linear and quadratic terms with (i) open triangles
 displaying previously published data and (ii) new data reported in this paper as solid dots.  Data has not been 
weighted.}
\label{fig3}
\end{figure}
\begin{figure}[htbp]
\centering
{\includegraphics[width=\linewidth]{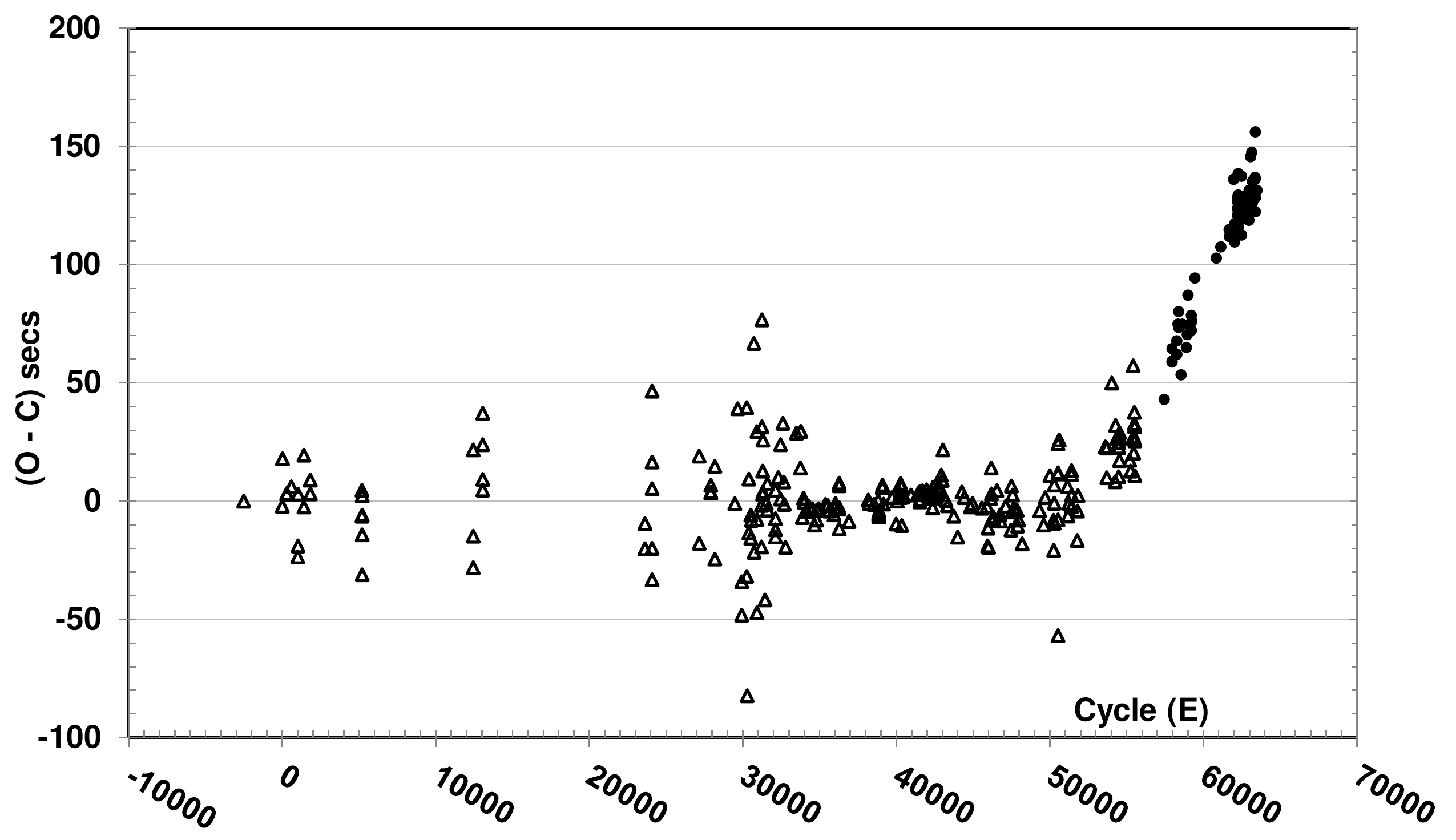}}
\caption{(O - C) residuals for HS0705+6700 after removing the linear, quadratic and sinusoid terms with 
(i) open triangles displaying previous published data and (ii) new data reported in this paper as solid dots.  
Data has not been weighted.}
\label{fig4}
\end{figure}
\\
\noindent The cyclical fit pre E = 55000 is thought by many to be compelling evidence of a circumbinary object. Recent observations suggest that this system is entering a phase where (O – C) residuals are offset by $\sim$40s from the original ephemeris.  This change is indicative of the original circumbinary hypothesis being no longer applicable.  More data is needed to see whether new circumbinary parameters can be computed for this system. An Applegate type magnetic effect might provide an alternative explanation for the observed light travel time effects (LTT). 

\noindent\\
These results undermine the earlier third body hypothesis. To a large extent, the original circumbinary hypothesis is based upon a dearth of data between 2002 and 2007 with data pre-2009 showing a degree of scatter (1$\sigma$ $\sim$25s) very much larger than would be expected from a third body alone. This latter point remains unresolved.
\subsection{HS2231+2441}
\subsubsection{Background:}
The discovery paper for HS2231+2441 was published by \cite{Ost_a} and described as a 14.5 magnitude short period, $\sim$ 2.7hr, detached eclipsing binary system and thought to consist of a post main sequence sdB star with a low-mass dwarf M companion. Their analysis assumed a canonical mass of the primary of 0.47$M{_\odot}$, so helium burning.   From spectroscopic measurements \cite{Ost_b} noted a significant inconsistency between the mass, radius and surface gravity of the primary and revised down the primary mass from its canonical value to  0.265$M{_\odot}$ with a radius of  0.164$R{_\odot}$ thus placing the primary as a post RGB object rather than an assumed core helium burning extreme horizontal branch star.  With a mass ratio of 0.16 the secondary becomes a substellar companion. \cite{Alm_b} published a conference paper suggesting that the primary component was of an even lower mass,  0.190$M{_\odot}$, but notably recording the semi amplitude radial velocity of the primary as 37.8km/s, some 20\% lower than that recorded by \cite{Ost_a}. A final confirmation of the true nature of this system would require the spectral signature originating from the secondary component.\\
\noindent\\
\cite{Qian_b} described observations of the system’s period between 2005 and 2009, and fitted a quadratic and a sinusoidal term to the (O – C) residuals.  They suggested the decrease in orbital period is associated with magnetic braking and a tertiary companion, with orbital period of 4.76 years, being responsible for the “apparent” cyclical variation.  Using the absolute parameters from \cite{Ost_a} they calculated the minimum mass of the tertiary companion as 0.0133$M{_\odot}$.\\  
\noindent\\
Although neither Qian nor \O stensen have published their times of minima, \cite{Lohr} extracted \cite{Qian_b}'s timings from their Fig. 4.  They also derived eclipse timings for this system from archival SuperWASP data.  \cite{Lohr} noted that the SuperWASP data fitted well Qian’s circumbinary hypothesis during 2006 and 2007, but did not strongly support the fit outside this date range, suggesting that a linear function might provide a better fit to the full data set.
\subsubsection{Recent data and ephemeris:}
We have observed 26 primary minima and three secondary minima between 2014 December and 2017 January and these are shown in the residuals plot, Fig. 5, where we have reproduced Fig.4 from \cite{Qian_b}, but with an extended timebase.  This diagram shows their unpublished data which we recovered by de-constructing their (O – C) diagram, the SuperWASP data from \cite{Lohr} grouped into 30 day bins and our recent data.  The uncertainties introduced through de-constructing the \cite{Qian_b} data is estimated to be $\pm$ 70 periods and $\pm$0.00005 days.\\

\begin{figure}[htbp]
\centering
{\includegraphics[width=\linewidth]{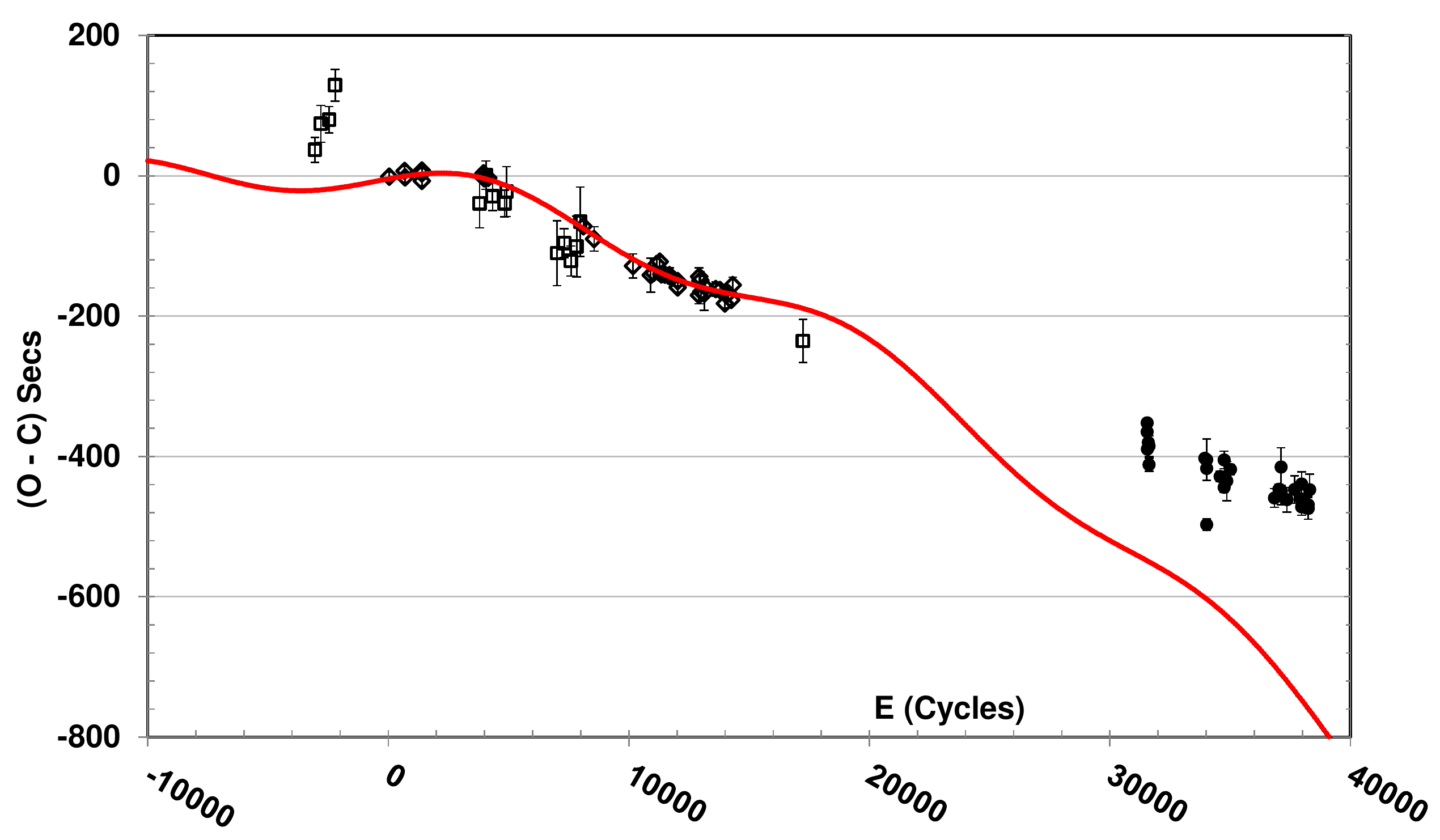}}
\caption{Residuals for HS2231+2441 using Ostensen's linear ephemeris.   Black  open diamonds is the deconstructed
 data from Qian et al. (2010); the open squares are data from SuperWASP and our latest data shown as solid 
circles.  The solid line shows Qian's quadratic ephemeris with cyclical behaviour of a possible circumbinary object.}
\label{fig5}
\end{figure}
\noindent
As noted by \cite{Lohr}, their early and late SuperWASP data is not in good agreement with \cite{Qian_b} circumbinary hypothesis suggesting that a linear ephemeris may be more appropriate.  Our new data, with minima occurring some 150s before Qian’s predictions, support the conclusions of \cite{Lohr}.\\
\noindent\\
Using the de-constructed data from \cite{Qian_b} and our new data we have calculated both a linear and quadratic ephemeris over a time span of 12 years and find:
\begin{equation}
\hspace{0.25cm}T_{min,BJD}=2453522.66963(6)  + 0.110587855(2) * E
\end{equation}
\begin{equation}
\begin{split}
\hspace{0.25cm}T_{min,BJD}= 2453522.66980(9)+0.110587829(11)*E\\+ 6.46(3)*10^{-13}*E^2 
\end{split}
\end{equation}
\noindent
The Mann-Whitney U test was used to compare the (O – C) residuals of the linear ephemeris with those of the quadratic ephemeris and find the null hypothesis is upheld.  We conclude that there is no significant difference in the magnitudes of the two sets of residuals and thus no preference for either ephemeris.  The calculated RMS of the weighted residuals also showed similar values of 2.47 and 2.37 for the linear and quadratic ephemerides respectively.  Thus we have opted for the simpler linear ephemeris where the plot of the residuals is shown in Fig. 6 and where we have added back in the SuperWASP data. Whilst the linear fit looks appropriate, the data scatter appears comparable to the measurement uncertainties in the data and could indicate the presence of other factors.
\begin{figure}[htbp]
\centering
{\includegraphics[width=\linewidth]{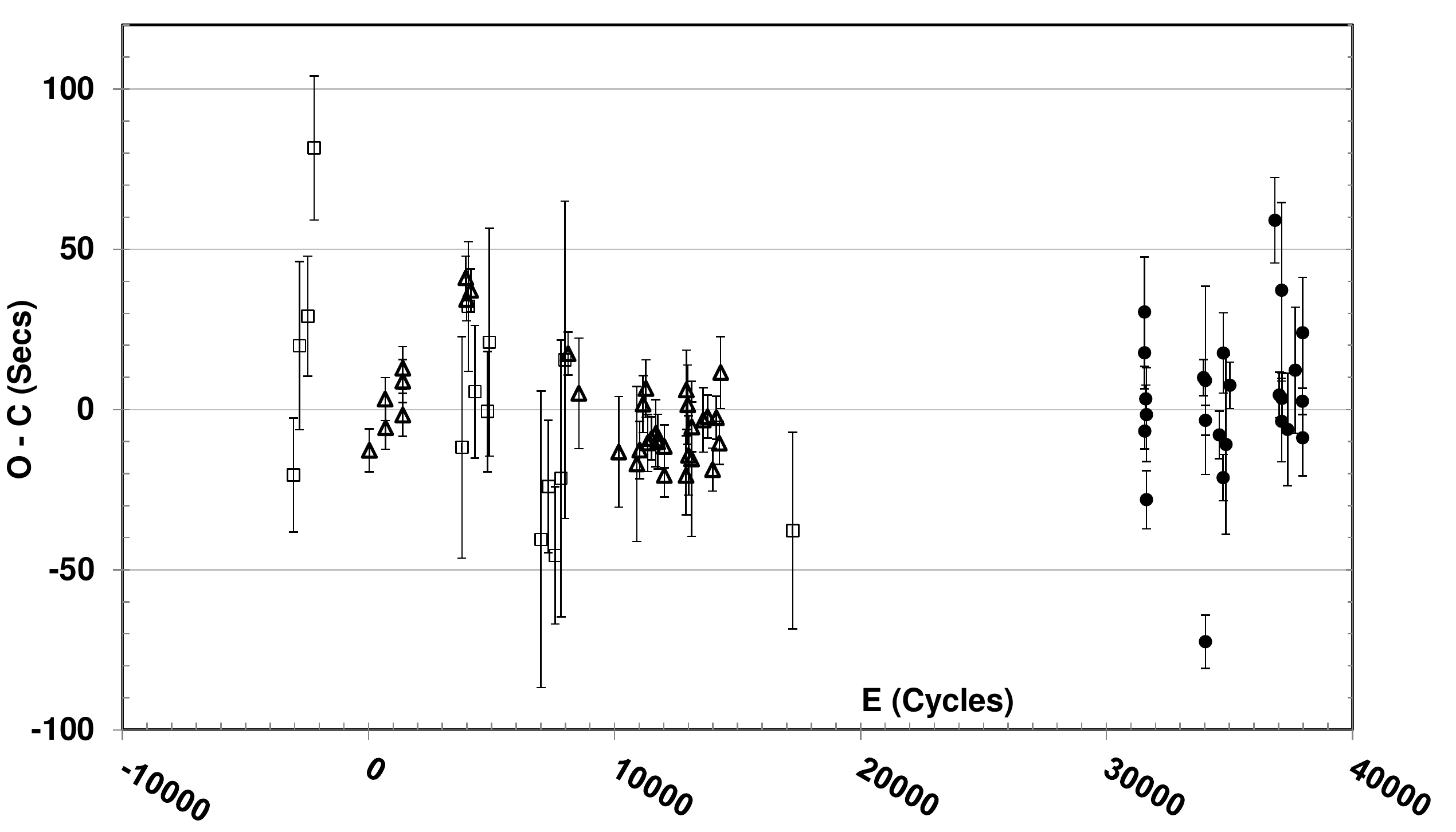}}
\caption{The (O - C) residuals for HS2231+2441 derived from the new linear ephemeris of Eq. 6.  
The deconstructed Qian data is shown as open triangles; our new data is shown as solid circles.  The 
binned SuperWASP data is shown as open squares.
}
\label{fig6}
\end{figure}
\subsection{J082053.53+000843.4}
\subsubsection{Background:}
J082053.53+000843.4, for brevity J08205, is a 15.2 magnitude short period detached eclipsing binary system. Like others in this paper, J08205 belongs to a subclass of the HW Vir family of very short period, $\sim$2.3 hrs, eclipsing binary systems comprising a post main sequence hot subdwarf star (O or B-type) and most probably a brown dwarf companion.\\
\noindent\\
J08205 was identified as an eclipsing sdB binary by \cite{Geier_b} as part of the Massive Unseen Companions to Hot Faint Underluminous Stars from SDSS (MUCHFUSS) project.  Further publications by \cite{Schaff} and \cite{Geier_c} focused on identifying the system parameters from both spectroscopic and light curve analyses. \\
\noindent\\
Prior to 2016 April there had been no published eclipse timings for this system and, consequently, no indication as to whether there was any variation in the binary period.  \cite{Pull_a} provided the first comprehensive set of eclipse timings for this system and laid the foundation for exploring the presence of a potential third body in a system where magnetic quadrupole effects generated by the companion brown dwarf are expected to be much reduced or non-existent.\\
\subsubsection{Recent data and ephemeris:}
We recorded 19 sets of primary minima and 6 secondary minima collected between 2014 December and 2015 February.  Details of this and our analysis were published in \cite{Pull_a}.  Subsequently we have recorded a further 14 primary minima between 2015 November and 2017 February.  These times of minima, together with the 20 earlier primary minima, allowed the computation of a system ephemeris spanning 26 months of recorded data.  We refined this further with the inclusion of an unpublished minimum from Schaffenroth extending the measurement timebase back to 2012 January.  This improved the precision of the calculated binary period by two orders of magnitude.\\\\
We explored both a linear and quadratic ephemeris and found that the Mann-Whitney U test identified no significant difference in the magnitude of the residuals between these two ephemerides.  Likewise there was no significant difference in values of RMS weighted residuals of 1.127 and 1.131 and so we have opted for a new linear ephemeris:
\begin{equation}
\hspace{0.25cm} T_{min,BJD}=2457016.61105 (2) + 0.096240737(7) * E 
\end{equation}

\begin{figure}[htbp]
\centering
{\includegraphics[width=\linewidth]{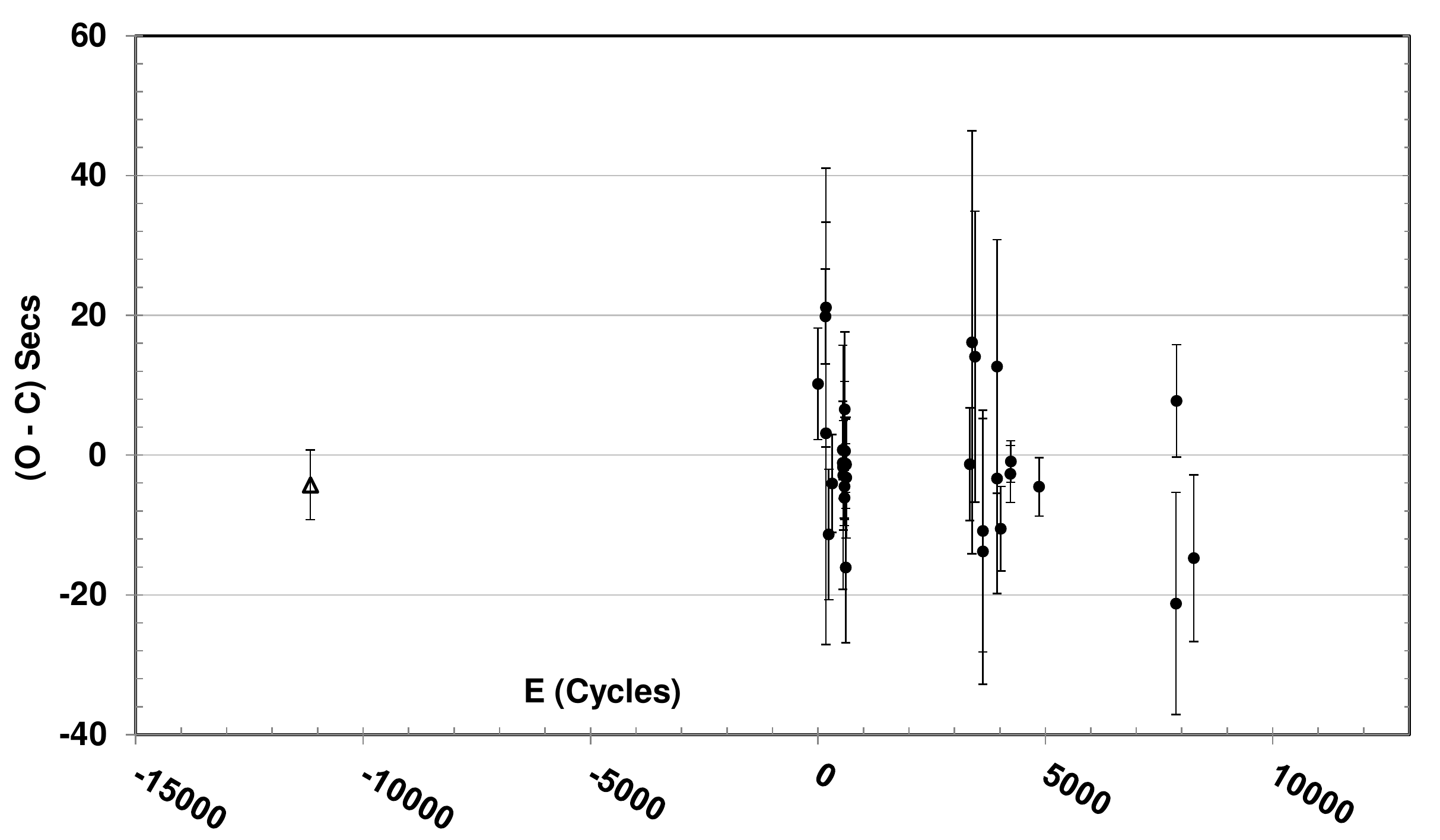}}
\caption{(O – C) residuals for J08205+0008 with our new data grouped between 3000<E<8000.  The 
single open triangle represents Schaffenroth's single data point of 2012 January.}
\label{fig7}
\end{figure}

\noindent This ephemeris has been used to calculate the (O – C) residuals plotted in Fig. 7. From this analysis there is no clear evidence for a systematic change in binary period. If quasi-cyclical light travel time effects are present then they are most probably restricted to a semi-amplitude of no greater than 20 seconds. Whilst these results suggest the binary period is unchanging, confirmation will require further observations.
\subsection{NSVS 07826147}
\subsubsection{Background:}
NSVS 07826147 was first identified as a short period, $\sim$3.9 hrs, eclipsing sdB binary with an  M2 or later main sequence companion by \cite{Kelly}. Spectroscopic analysis by \cite{For} suggested that the sdB had a mass lower than its canonical value together with an eclipsing M5 companion.  They fitted a linear ephemeris to 21 times of minima observed between 2008 February and 2009 March.  \cite{Zhu_c} produced a further 16 times of minima between 2009 March and 2009 August.  Their revised period for this system fell within the measurement uncertainty of that predicted by \cite{For}.  \cite{Back} determined a further seven times of minima between 2011 February and 2011 October and included a 2005 February observation from \cite{Drake} but noted that the uncertainty of this result is two orders of magnitude greater than their uncertainties.  As a consequence we have not used Drake’s result in our analysis. \cite{Lohr} provided additional timings from SuperWASP extending the time line back to 2004 May.\\
\noindent\\
\cite{Zhu_a} were first to claim detecting a cyclical change in the period of this system, further confirmed by \cite{Zhu_b} reporting that this periodic change could be caused by a circumbinary object of mass greater than 4.7$M_{J}$ with an orbital radius of 0.64 AU introducing a light travel time effect of $\sim$0.00004 days ($\sim$3.5s).  They do not state a period but their (O – C) diagram suggests $\sim$11000 cycles, equivalent to $\sim$4.9 years.
\subsubsection{Recent data and ephemeris:}
We report a further 20 times of primary minima observed between 2015 April and 2017 June. These new data points together with the 35 previously published observed times of primary minima extend the time line to over 13 years.  We explored both a linear and quadratic ephemeris and found that the Mann-Whitney U test identified no significant difference in the magnitude of the residuals between the two ephemerides.  Likewise there was no difference in values of RMS weighted residuals and so opted for a new linear ephemeris:
\begin{equation}
{ \hspace{1cm} T_{min,BJD}=2455611.92655(1) + 0.161770449 (2) * E }
\end{equation}
\begin{figure}[h]
\centering
{\includegraphics[width=\linewidth]{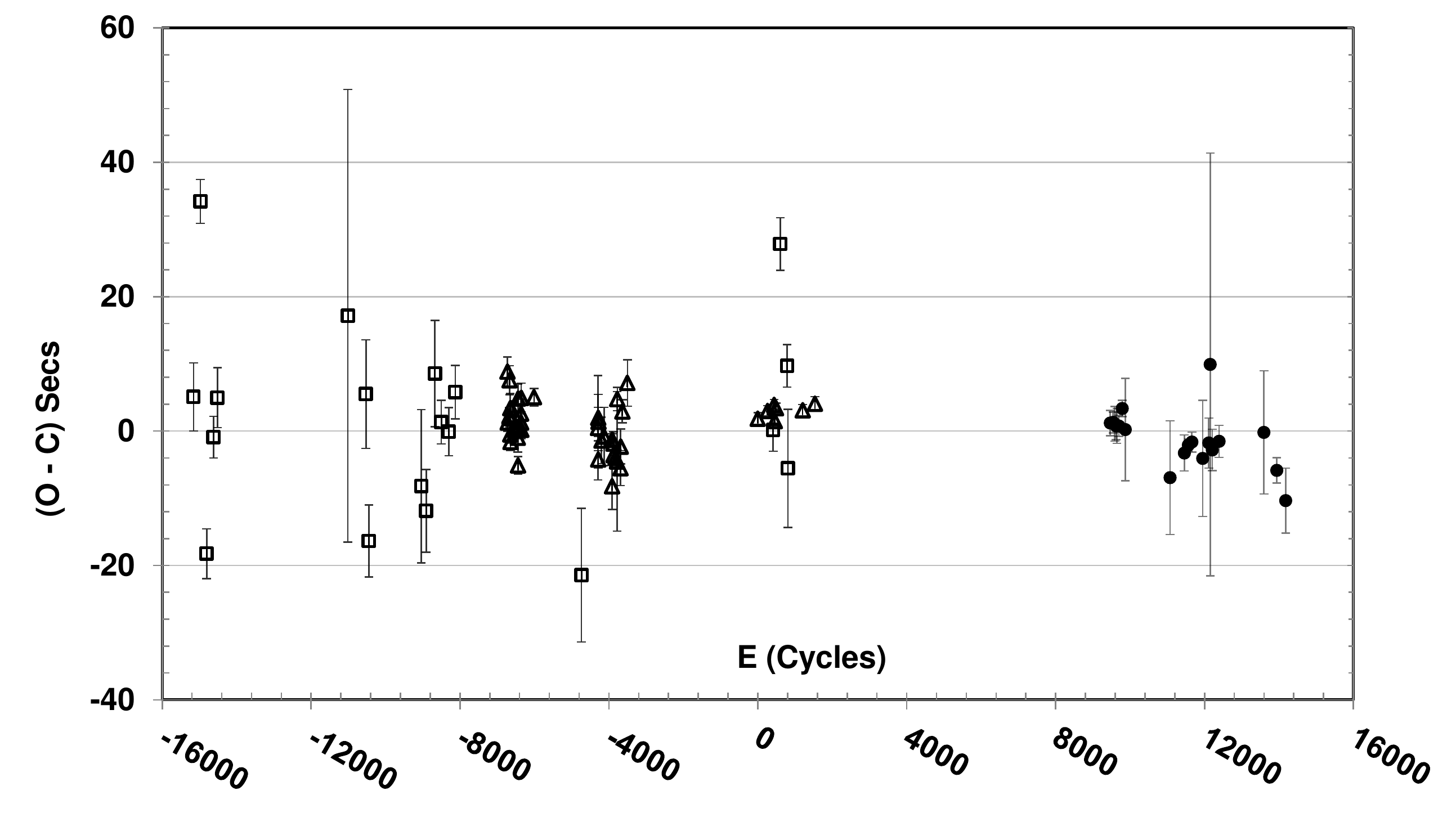}}
\caption{The (O - C) residuals for NSVS 07826147.  The historical data from For, Zhu and Backhaus shown as open triangles; SuperWASP binned data as open squares and our new data represented as solid circles.}
\label{fig8}
\end{figure} \\This ephemeris is similar to that of \cite{Back}.  With the new linear ephemeris, the computed (O - C) residuals are shown in Fig 8 where the diagram also includes the 19 binned data sets from SuperWASP, \cite{Lohr}, and nine observed secondary minima recorded by \cite{For}. The RMS of the weighted residuals was calculated to be 2.41.\\
\noindent\\
The plot of (O – C) residuals shows some degree of scatter with a $\sigma\sim$10s but there is no strong indication of any long-term change in the binary period.  \cite{Zhu_b} indicated the presence of a circumbinary object with an LTT of $\sim$3.5s which is much less than the large scatter in the residuals ($\sim$10s). As yet they have not published any new data or analysis supporting their hypothesis so that drawing such conclusions may be premature.
\subsection{NSVS 14256825}
\subsubsection{Background:}
  NSVS 14256825 was discovered as a 13.2 magnitude variable star in the NSVS survey, \cite{Woz}.  \cite{Wils} identified this object as an eclipsing binary system with an sdB primary component having an orbital period of $\sim$2.6 hr.  From their measurements, they produced V, B and Ic light curves and documented 21 primary eclipses over a three-month period.  They determined an ephemeris and suggested this object was similar in many respects to both the prototype sdB/M-dwarf system HW Vir and HS0705+6700.\\
\noindent\\ 
\cite{Qian_b} suggested that the binary period of NSVS 14256825 had a cyclical change, but did not provide any supporting evidence.  \cite{Kilk_d} provided a further nine times of minima and showed that the orbital period was increasing at a rate of ${\sim}1.2 x 10^{-12}$ days per orbit. \cite{Beu} presented additional eclipse timings and their analysis of the (O – C) residuals suggested the presence of a circumbinary planet of  ${\sim}12M_{J}$ but with an uncertain period and large eccentricity. \cite{Alm_c} provided light curve and spectroscopic analysis whilst comparing NSVS 14256825 with nine other sdB binary systems.\\
\noindent\\
\cite{Alm_a} presented ten new eclipse times obtained between 2010 July and 2012 August. They also combined their data with previous published measurements and performed a new orbital period analysis. They interpreted the observed eclipse timing variations as being the result of light travel time effects introduced by two circumbinary planets with orbital periods of 3.5 and 6.9 years and masses $3M_{J}$ and $8M_{J}$ respectively. \\
\noindent\\
\cite{Hinse} questioned the validity of these results. While searching extensively for a minimum in residuals, their approach predicted the orbital parameters and minimum mass of a single circumbinary companion.  They performed two independent analyses based on two datasets (i) Dataset I as presented in Table 3 of \cite{Alm_a} spanning an observing baseline of around 5 years and (ii) Dataset II as Dataset I but with three earlier primary eclipse data points from \cite{Beu} extending the observing baseline to some 13 years, so more than doubling the time window.  With Dataset I they derived parameters similar to Almeida but concluded that this was a local minimum and a much longer baseline of data was needed to confirm this hypothesis.  They found no evidence for a second companion.  From Dataset II they derived a much longer cyclical period of 102 years for a circumbinary object but again concluded that the baseline was inadequate to draw firm conclusions. From their extensive statistical analysis they showed that the data did not constrain any particular model with well-established confidence limits. \\\\
\cite{Witt} also challenged the two planet hypothesis from the standpoint of orbital stability where they concluded that the planetary system would have a life expectancy of less than 1000 years.\\
\noindent\\
More recently \cite{Nasi} provided 83 new eclipse timings spanning the period 2009 August to 2016 November.  Including earlier data, they analysed three data sets and reported that their Dataset C, which excluded data from NSVS, All Sky Automatic Survey (ASAS) and SuperWASP, indicated a possible brown dwarf circumbinary companion.

\subsubsection{Recent data and ephemeris:}
We have added 19 new primary minima observed between 2015 August and 2017 July.  Our new data, together with that from \cite{Nasi} allowed us to test the circumbinary stellar object hypothesis of \cite{Alm_a}. We did not test the claims made by \cite{Qian_b}, who have yet to publish their parameters, nor have we tested the findings of \cite{Beu} and \cite{Hinse}, where both recognised that their derived circumbinary parameters were poorly constrained. \\
\noindent\\
To investigate further the \cite{Alm_a} hypothesis we use their ephemeris.
\begin{equation}
\hspace{0.25cm}T_{min,BJD}=2454274.20874(4) + 0.1103741681(5) * E 
\end{equation}
From this linear ephemeris and observational data we calculate the RMS of the weighted residuals to be 20.95.
\begin{figure}[h]
\centering
{\includegraphics[width=\linewidth]{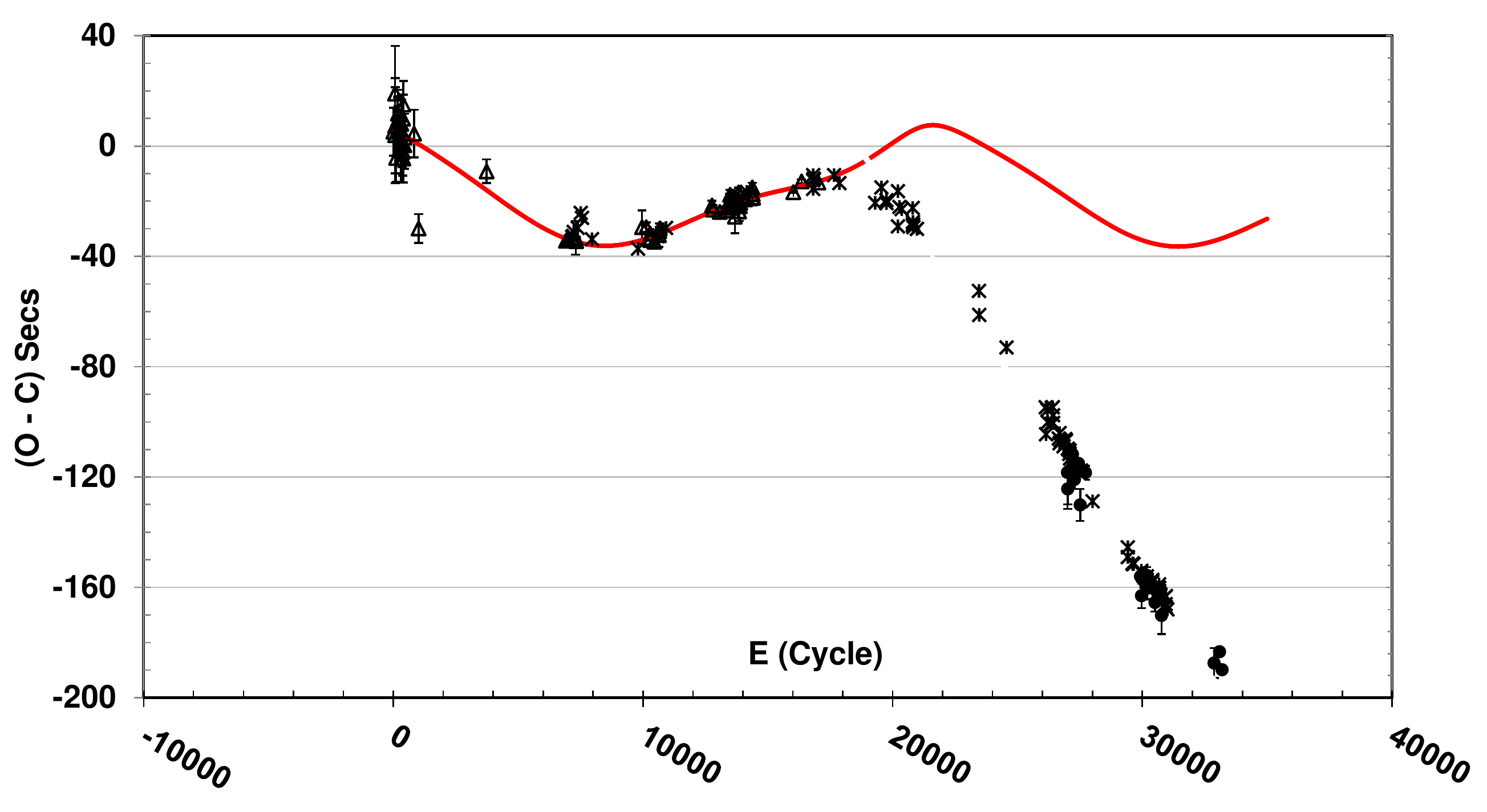}}
\caption{Residuals for NSVS 14256825 calculated from the ephemeris and orbital data from Almeida (2013). Open triangles represent data from Almeida;  stars represent data from  Nasiroglu (for clarity error bars not shown); our new data is shown as solid circles.  The solid line represents Almeida's two planet prediction showing departure at E $\sim$ 18900
.}
\label{fig9}
\end{figure}
\begin{figure}[h]
\centering
{\includegraphics[width=\linewidth]{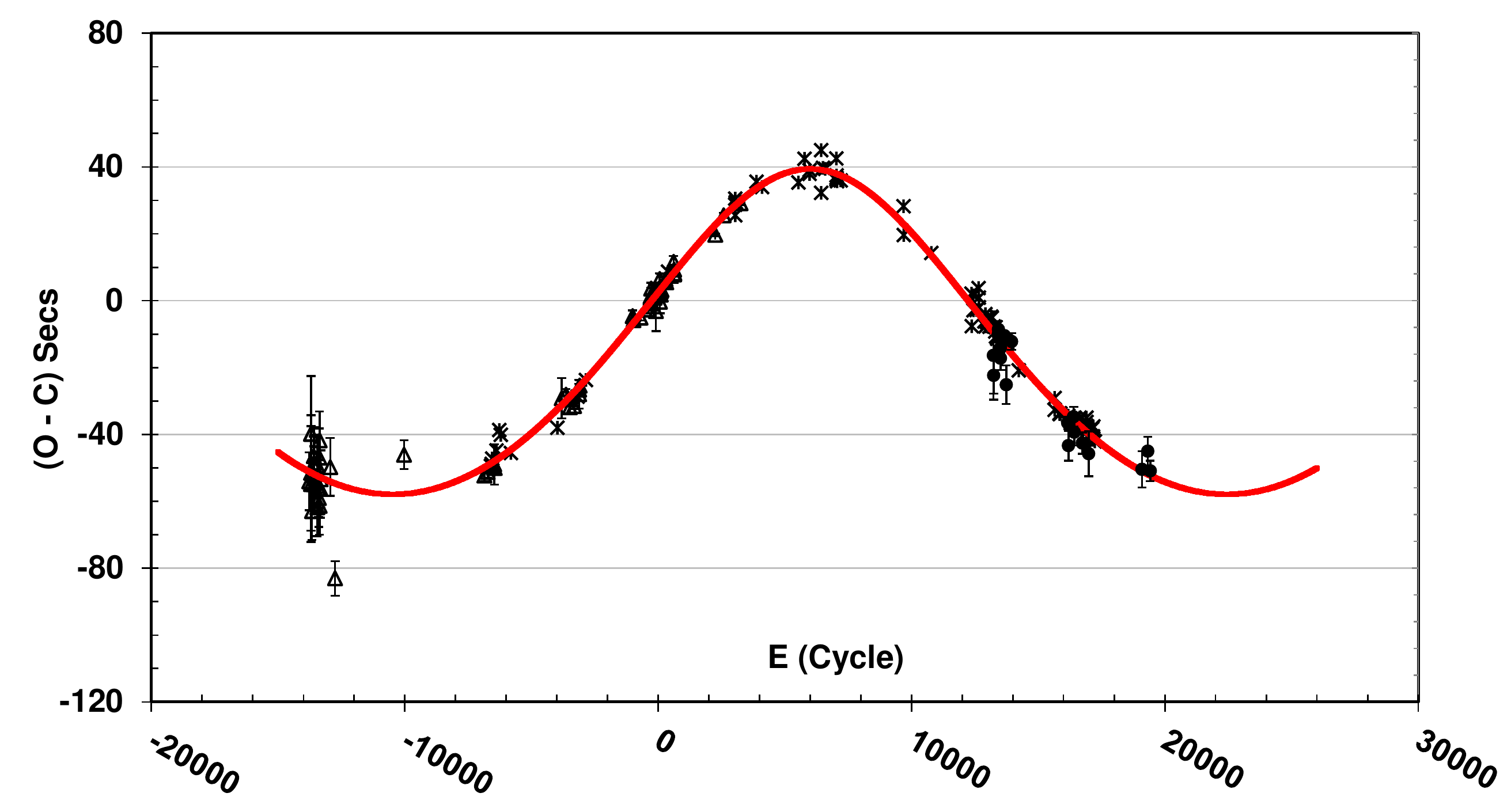}}
\caption{Residuals for NSVS 14256825 calculated from the ephemeris and orbital data from Nasiroglu (2017). Open triangles represent data from Almeida;  the stars represents data from Nasiroglu; our new data is shown as solid circles.  The solid line represents Nasiroglu's circumbinary brown dwarf hypothesis. 
}
\label{fig10}
\end{figure}
\hspace{12pt}Using our new data and all available historic data we have plotted the (O – C) residuals shown in Fig. 9 and have included predictions of Almeida’s two planet hypothesis.  We find that from 2013 March (E $\sim$ 18900) calculated residuals depart significantly from Almeida’s predictions increasing to 160s at E $\sim$ 30,000.
\\\\
\noindent\cite{Nasi} proposed an the alternative solution of a single circumbinary brown dwarf companion of mass 14MJ with a 9.7 year orbital period and 0.2 eccentricity.   However they noted that their new parameters ”...substantially differ from previous estimates.”  Using their linear ephemeris, Eq. 11, we find our new data is consistent with their Dataset C as shown in Fig. 10.
\begin{equation}
\begin{split}
\hspace{0.25cm}T_{min,BJD}= 2455793.84005(3) + 0.110374099(3) * E  
\end{split}
\end{equation}
\noindent
\\\\
We tested statistically the fit between the (O – C) residuals and Nasiroglu’s single circumbinary brown dwarf hypothesis and found the fit is significant at the 3$\sigma$ level.  We conclude that the current data supports the Nasiroglu hypothesis but caution that the data only spans one circumbinary cycle and confidence in this model will require a significantly longer timebase.
\subsection{NY Vir}
\subsubsection{Background:}
From spectroscopic and light curve analysis, \cite{Kilk_b} first reported NY Vir (PG1336-018) as a detached short period, $\sim$ 2.4hr, eclipsing binary consisting of a pulsating hot sdB star and a probable dwarf M5 companion.  From their observations, including 10 times of minima observed between 1996 May and 1996 July, they computed a linear ephemeris with binary period of 0.1010174 days. \cite{Kilk_c} published a further 12 times of minima extending the time line to 2010 June and noted that the eclipse times departed significantly from those predicted by a linear ephemeris. With their new data, they showed that the period was rapidly changing, fitting a quadratic ephemeris to the (O - C) residuals and noting that, due to the sdB pulsations, the eclipse timings were less accurate than would normally be expected.  \cite{Cam} found a similar effect, publishing a further 7 times of minima between 2009 May and 2011 June and suggesting a possible light travel time effect caused by the presence of a third body.\\
\noindent\\
A further 9 times of minima were added by \cite{Qian_c} who favoured a quadratic ephemeris with a third body contributing a light travel time effect of amplitude 6.3s, a period of 7.9 years and mass greater than $2M_{J}$.  They also suggested that the rapid period reduction could be part of a long-term effect from a fourth body.  \cite{Lee} provided a further 39 times of minima between 2011 January and 2014 May.  They also included four earlier weighted mean datasets from \cite{vucko} and their analysis agreed well with that of \cite{Qian_c} indicating a third body of period 8.18 years, a mass of $2.78M_{J}$ and a light travel time effect of 6.9s.  They also postulated a fourth planetary body of period 27.0 years, mass of $4.49M_{J}$ and light travel time effect of 27.3s to explain the rapid period reduction.  However, with a predicted orbital eccentricity of zero the orbital stability of this four-body system was calculated to be only 800,000 years.  As a consequence they suggested that a longer timebase was required to substantiate the presence of the fourth body and that it should have a moderate eccentricity to bring long-term orbital stability.\\
\subsubsection{Recent data and ephemeris: }
We added a further 15 times of primary minima observed between 2015 April and 2017 June extending the timebase a further 3 years.  In addition, we have added 4 times of minima extracted from the database of the AAVSO of which two data points fall within the latter part of the data range of \cite{Lee} and two within the range of our new data.  We employed the same criteria for data selection as used by Lee taking only primary minima that had uncertainties of less than 0.0001 days.  We investigated the secondary minima but found they gave a large degree of scatter about the predicted trend line.  This was most probably due to the amplitude of the pulsations being comparable to the depth of the secondary minima; see for example \cite{Kilk_b} Fig.3. \\ 

To enable us to make a direct comparison with Lee et al. (2014) two planet prediction we used their linear ephemeris.
\begin{equation}
\hspace{0.25cm}T_{min,BJD}=2453174.442699(91) + 0.1010159668(43) * E
\end{equation}
The calculated RMS of the weighted residuals was found to be 36.45. Examining the data we find that the uncertainties associated with Vuckovic’s two primary minima were very much lower than the remaining 75 uncertainties.  These two uncertainties were found to be significantly biasing the RMS value and, after removing them, the RMS of weighted residuals decreased to 4.78. 
\begin{figure}[h]
\centering
{\includegraphics[width=\linewidth]{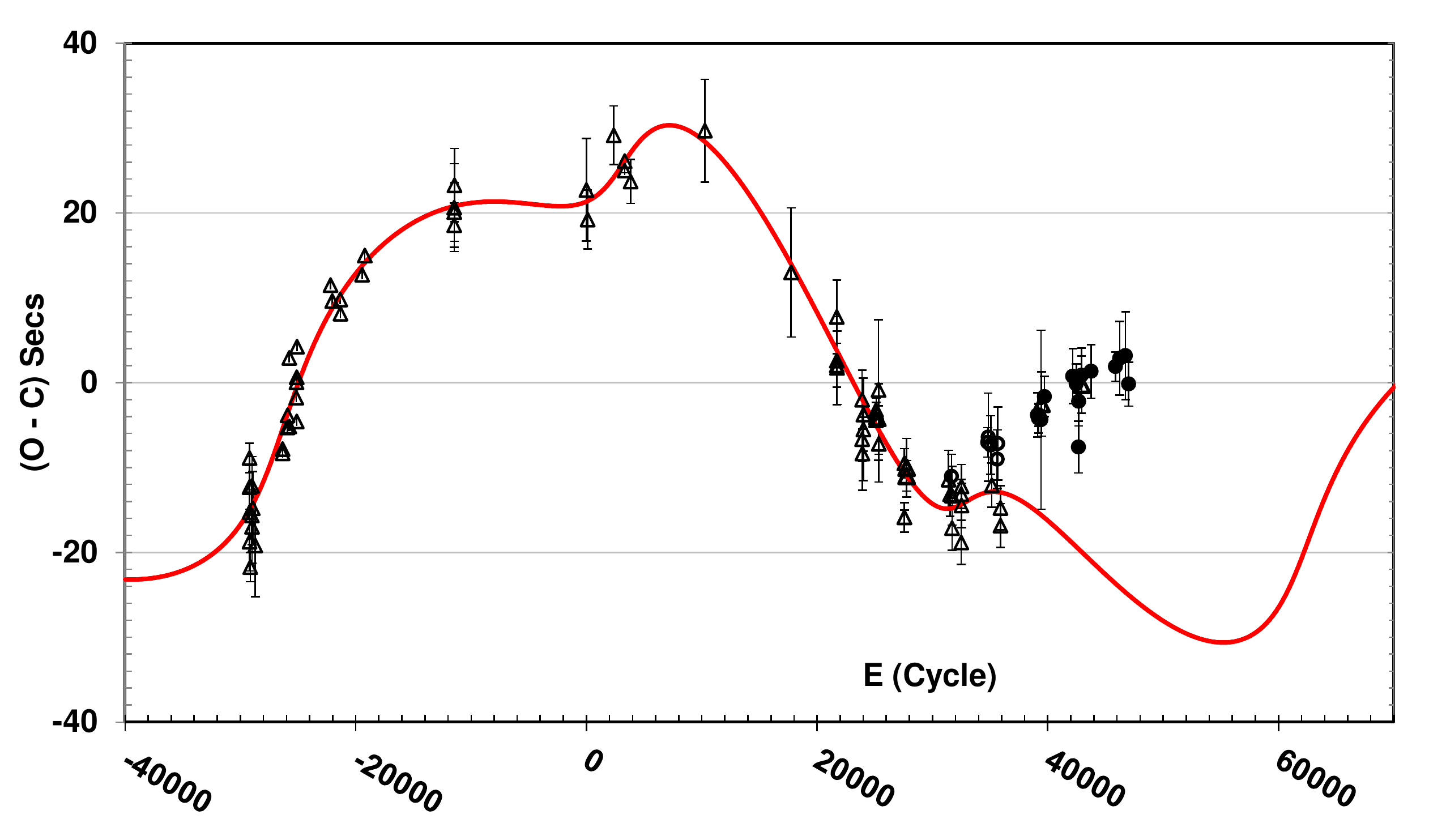}}
\caption{NY Vir residuals using Lee's linear ephemeris plus two planet hypothesis.  Lee's data is shown as open triangles (E $\sim$< 35000); our new data shown as solid circles; Kilkenny's 2014 data is shown as open circles
(E $\sim$ 33000)  and data from  AAVSO as open triangles (E $\sim$ 28000 and 40000).  Lee's two planet hypothesis is 
shown as the solid line.}
\label{fig11}
\end{figure}
Fig. 11 shows our new data points relative to Lee's two planet hypothesis. Our new data show that the eclipses occur between 10 and 20s before those anticipated by Lee's model whilst the four AAVSO data points (E = 27769, 27770, 39449 and 42934) tie together our new data with the prediction from Lee.  \cite{Kilk_e} also indicates a departure from Lee’s predictions. 

\section{Discussion}
\subsection{sdB binaries: 20 years of eclipse time data }
The last decade has seen many predictions made for the existence of circumbinary companions orbiting post common envelope eclipsing sdB binary systems.  It was found that these objects were typically greater than several Jupiter masses and with orbital periods of between 3 to 30 years. We have reviewed seven of these systems, of which five (HS0705+6700, HS2231+2441, NSVS 07826147, NSVS 14256825 and NY Vir) have previously been reported as most probably having circumbinary companions. Of the remaining two systems, EC10246-2707 has 20 years of observational data but shows no clear indication of cyclical light travel time effects. The second system, J08205+0008, has five years of data and though there is little evidence for LTT effects, it would be premature to draw any firm conclusions about the presence, or not, of a circumbinary companion.\\
\noindent\\
Of the five systems reported as possibly having circumbinary companions three, HS0705+6700, NSVS 14256825 and NY Vir, have recently shown significant changes to their predicted eclipse times.  These changes might indicate a longer time baseline is needed to validate the circumbinary hypothesis but could also suggest that other mechanisms, e.g. magnetic effects, are driving the observed eclipse time variations.\\
\noindent\\
In early 2013 NY Vir began to show a departure from the predictions of \cite{Lee}. These changes were noted by both \cite{Kilk_e} and ourselves (\cite{Pull_b}. Our new data does not rule out the possibility of circumbinary planets but does raise questions as to whether the 21 years of observations is sufficient to base predictions on.\\
\noindent\\
NSVS 14256825 also provides conflicting evidence for the circumbinary hypothesis.  \cite{Alm_a}) proposed a two circumbinary solution with periods of 3.5 and 6.7 years respectively which has been challenged by \cite{Hinse} and others  and latest observations no longer support the Almeida model. Single circumbinary solutions have been put forward by \cite{Beu} suggesting a broad range of possible periods extending from 14 to 30 years and \cite{Hinse} who preferred a period of either 2.5 or 6 years, depending upon the dataset used. \cite{Qian_a} also recognised a periodicity in the data but they have not yet specified a value for the period.  In their review of this object, \cite{Nasi} computed new circumbinary parameters and remaining open minded about its presence; they predicted a possible circumbinary brown dwarf with a period of 9.7 years. Future observations will either support the circumbinary hypothesis or provide evidence against it. In any case, longer baselines covering full orbits of potential circumbinary objects are clearly needed. \\
\noindent\\
HS0705+6700 provides yet another example of conflicting evidence for the circumbinary companion hypothesis. Data covering more than 14 years has shown quasi periodicity in its eclipse timings with most observers agreeing on a circumbinary period of between 7.2 and 8.9 years.  Whilst providing nearly two orbits of data, and as confidence grew in this model, in late 2015 this system ceased exhibiting its quasi cyclical LTT behaviour and now shows an apparent constant period. \\
\noindent\\
Of the remaining two systems with circumbinary companion claims, HS2231+2441 is somewhat different. Whilst one investigator suggested a circumbinary object of period 4.76 years we find that \cite{Lohr}’s suggestion of a constant period and no significant LTT effects as more consistent with the longer term observations. Our new data extends the timebase to some 12 years and we find little convincing evidence of cyclical LTT effects and suggest a revised constant period of 0.110587855 days.\\
\noindent\\
Finally only one investigator, \cite{Zhu_b}, has detected a possible circumbinary companion to the sdB system NSVS 07826147.  Although they have published little on this, their diagrams suggest a period of about 4.9 years with an LTT amplitude of 3.5s. This amplitude is small, and could be considered within observational noise. Moreover, our new data observed through 2015 and 2016 provides no clear evidence to support their circumbinary claims. 
\subsection{A changing landscape}
While some authors have challenged specific details of the circumbinary planet hypotheses, others have challenged the very existence of such planets around close binaries.  \cite{Witt} challenged Almeida’s two planet hypotheses for the binary system NSVS 14256825 whilst \cite{horner2012} challenged the two planet model proposed for the HW Vir system.  Again, from an orbital stability standpoint, \cite{horner2014} challenged the two planet systems of three other binaries. \cite{hardy2015first}, using the new extreme-AO instrument SPHERE, imaged the prototype eclipsing post-common envelope binary V471 Tau in search of a brown dwarf that was predicted to be responsible for LTT effects. They found no direct evidence of such an object. More recently, \cite{bours2016long} carried out a long-term programme of eclipse time measurements on 67 white dwarfs in close binaries to detect period variations. They found that all systems with baselines exceeding 10 years, and with companions of spectral type M5 or earlier, appeared to show much larger eclipse timing variations than systems with companions of spectral types later than M5. They found this to be consistent with an Applegate-type mechanism. Nevertheless, they also considered it reasonable to assume that some planetary systems could exist around evolved white dwarf binaries, for example the NN Ser binary system (\citealp{beu2}; \citealp{marsh2013planets}; \citealp{Pars}; \citealp{Hardy}). 

\subsection{Statistical trends of sdB eclipse time variations}

Following the methodology of Bours et al. (2016), we investigated the correlation between the RMS residuals of the seven sdB binary systems reported here with their observational baselines and their spectral types.  Unfortunately the spectral types of the binary companions are somewhat uncertain as these stars are not spectroscopically visible and assessment has to be made indirectly through parameters that are generally poorly constrained.  Spectral types have not been published for HS0705+6700, and NSVS 14256825, but we have derived their values from their calculated secondary mass in conjunction with \cite{baraffe1996mass}.  Spectral types and RMS values are summarised in Table 2.\\

\begin{figure}[h]
\centering
{\includegraphics[width=\linewidth]{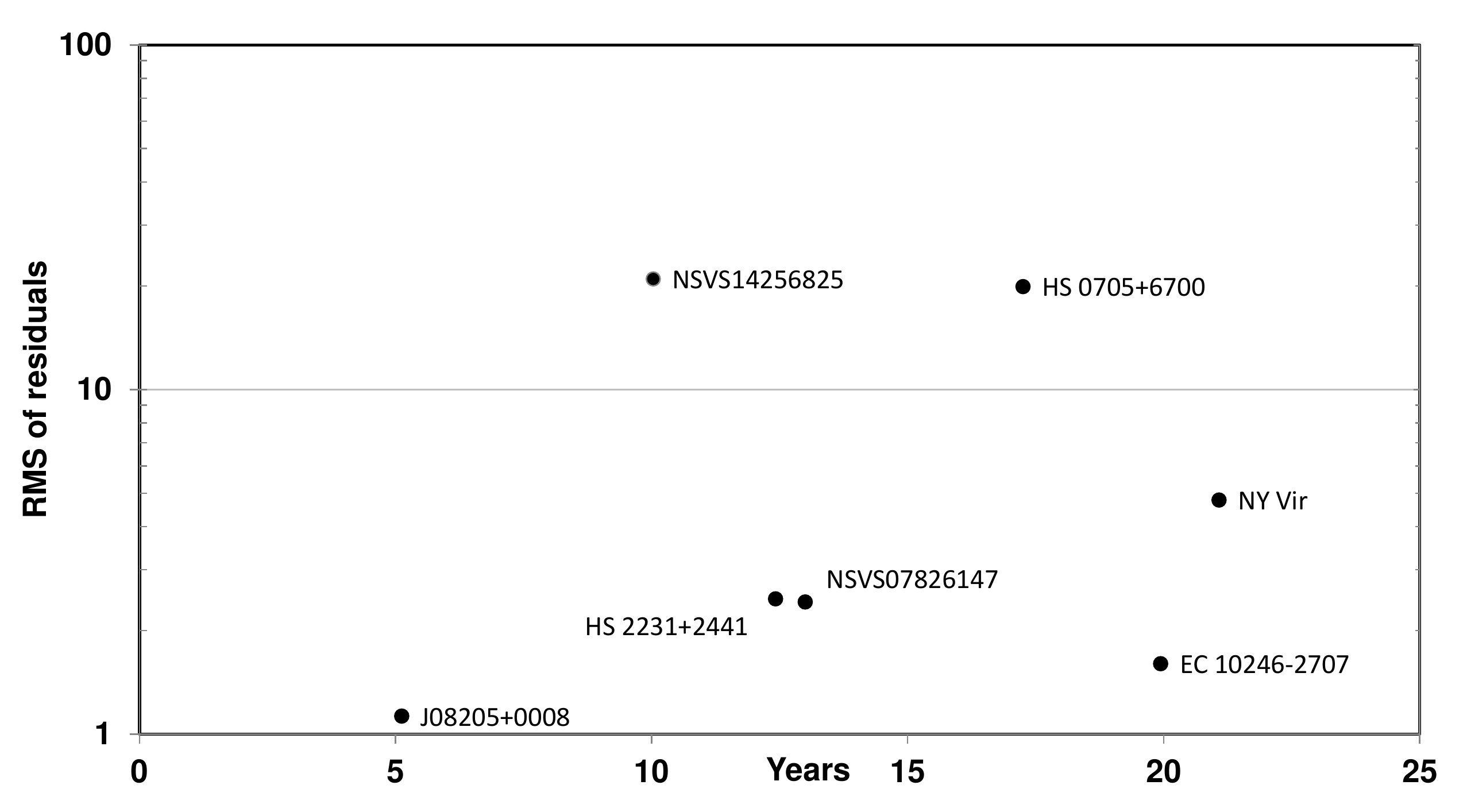}}
\caption{Calculated RMS of residuals as a function of observing period.  Only J08205+008
has an observing period of less than 10 years and as yet shows little LTT effect.  All of these systems fall short of the \cite{bours2016long} criteria of RMS $\sim$100.
}
\label{fig12}
\end{figure}
\noindent\\
Bours concluded that an observational baseline of at least 10 years was necessary to show RMS values saturating at around 100 and all white dwarf + M-dwarf binaries considered showed significant (O - C) residuals of the order of 100s.  Our observations are summarised in Fig. 12 where six of the seven systems have more than 10 years observing time but we find none of these systems approaching the RMS values of ~100.  Two systems, HS0705+6700 and NSVS14256825, showed strong LTT effects but with RMS values of only about 20.  We found four of the systems having an RMS value of less than 5 including one, NY Vir after removing the two Vuckovic data points, showing definite LTT effects.  Only HS0705+6700 showed large (O – C) residuals at the 100s level suggested by Bours. The remaining two systems, NSVS14256825 and NY Vir were very much less at 50s and 25s respectively.\\\\
\noindent
To achieve Bours RMS saturation figure of $\sim$100, Eq.3 suggests large cyclical LTT amplitudes (>100s) and/or small uncertainties (<0.00004 days) are necessary. The large values of RMS found by Bours could possibly be attributed to the observational equipment used which will tend to give lower values of uncertainties.  Historical observations of sdB binaries typically used much smaller telescopes (<1m) and less sophisticated cameras yielding typical uncertainties an order of magnitude, or more, larger.\\
\noindent\\
We also compared the RMS of weighted residuals with the spectral type of the sdB companion, see Fig. 13.  As noted earlier the methodology used to determine the spectral type of the secondary component results in values with a high degree of uncertainty.  Of particular note are EC10246-2707 and NSVS07826147 which both show a low RMS and (O – C) values but classified as M5 and potentially magnetically active.  This may indicate a spectral type misclassification or possibly both subdwarfs showing no magnetic activity.  The three systems exhibiting significant LTT effects, HS0705+6700, NSVS14256825 and NY Vir, appear to be possible M6 or earlier and in agreement with the findings of Bours.

\begin{figure}[h]
\centering
{\includegraphics[width=\linewidth]{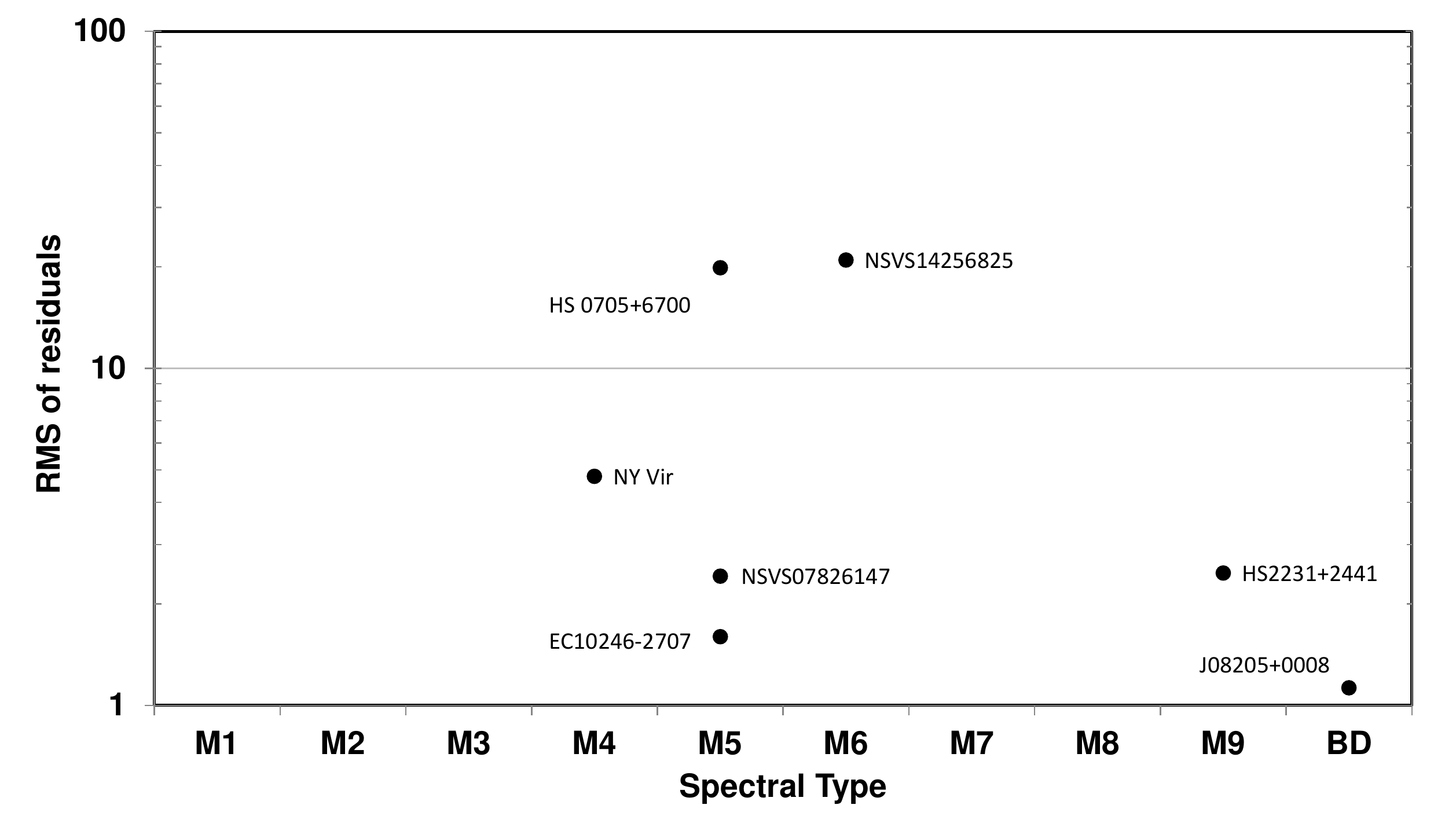}}
\caption{RMS of residuals with respect to spectral type.  There is some uncertainty of 
spectral type for all of these systems.  Of interest are EC10246-2707 and NSVS 07826147
classified as M5 but exhibiting little or no LTT effect over 15 years or more of observation.
}
\label{fig12}
\end{figure}
Whilst the sdB data is not in complete agreement with the Bours approach there remains the need to identify those systems which show real LTT effects from those that show eclipse time variations more likely caused by random measurement uncertainties.  Using an unweighted RMS statistic, i.e. setting sigma equal to unity in Eq.3, we can compare the mean 1$\sigma$ uncertainty levels of each system with their corresponding (O – C) residuals taken from a linear or quadratic ephemeris baseline.  These results are summarised in Table 4 where three systems, HS0705+6700, NSVS14256825 and NY Vir show greater than 3$\sigma$ variations due to LTT effects, indicating the possible presence of some physical phenomena e.g. circumbinary or magnetic effects.  Four systems do not meet the 3$\sigma$ criteria where two, EC10246-2707 and J08205+0005, show LTT effects at the $\sim$1$\sigma$ level suggesting there is unlikely to be any physical activity in these systems.  The remaining systems, HS2231+2441 and NSVS07826147, show LTT effects at the 2$\sigma$ to 2.5$\sigma$ level, possibly indicating some form of low level activity.  For comparison purposes, the Bours weighted RMS statistic is included in Table 4, which follows the same trend as our analysis.
 
 \begin{table}[htb]
\caption{Comparison of the LTT efect from a linear or quadratic baseline with 1$\sigma$ uncertainty levels.  The Bours' weighted statistic is shown for comparison where NY Vir shows two values, one for the exclusion and the other for the inclusion of Zorotovic's two ULTRACAM low uncertainty observations}
\label{4}
\resizebox{\columnwidth}{!}{%
\begin{tabular}{c*{6}{>{$}c<{$}}}
\hline\\
 \text{Object}        & \text{Measurement}       & \text{(O-C) Lin}       & \text{(O-C) Quad}       & \text{No of $\sigma$} & \text{No of $\sigma$} &\text{Bours}  \\
 \text{}        & \text{Uncertainty}       & \text{(days)}       & \text{(days)}       & \text{(Lin)} & \text{(Quad)} &\text{Weighted}  \\
 \text{}        & \text{days}       & \text{}       & \text{}       & \text{} & \text{} &\text{RMS}  \\
 \hline\\
\multicolumn{1}{l}{EC 10246-2707}  & 0.0000846 &0.0000983 & 0.0000800 & 1.16 & 0.95 & 1.60 \\
\multicolumn{1}{l}{HS0705+6700}  & 0.0001263& & 0.0007412 & & 5.87& 19.89 \\
\multicolumn{1}{l}{HS2231+2441}  &0.0001671&0.0004178&0.0004031&2.50&2.41&2.47\\
\multicolumn{1}{l}{J08205+0008}&0.0001256&0.0001177& 0.0001140&0.94&0.91&1.13\\
\multicolumn{1}{l}{NSVS07826147}&0.0000542&0.0001155&0.0001144&2.13&2.11&2.41\\
\multicolumn{1}{l}{NSVS14256825}&0.0000387&0.0003867& &10.0&&20.95\\
\multicolumn{1}{l}{NY Vir}&0.0000313&0.0001413&&4.51& &4.78/ \\
&&&&&&36.45\\
\hline
\end{tabular}
}
\end{table}

\subsection{Possible circumbinary companion transits}
It is conceivable that circumbinary companions may be detected transiting these binary systems especially if their orbital plane is closely aligned with the orbital plane of the binary system. The current predicted circumbinary companions have orbital periods measured in years and orbital distances in AU’s giving them a low probability of making a detectable transit. However, close-in circumbinary companions with orbital periods measured in days, and possibly in orbital resonance with the binary, might well make transits that could be detected in the system’s light curve. The expected impact parameter of such transits is expected to be low given that the orbital plane of the binary is closely aligned with the observer’s line of sight.  In general, the best eclipsing binaries to observe would be the less massive binaries. For such binaries transiting planets dim a greater fraction of their light than for more massive binaries. \\
\noindent\\
Whether such companions exist is speculative but there have been several occasions, and on different binary systems, where we recorded transit-like dimming in the systems light curve. First observed in 2013 December, the duration and depth of the dip looked very similar to a circumbinary transit.  More sightings will be needed before we can make and test predictions. \\

\section{Conclusions}

The complex energy interaction of detached short period sdB + Mdwarf/brown dwarf binary systems makes these systems difficult to model with best fit solutions often requiring the secondary component’s albedo to take on non-physical values of greater than unity, see for example \cite{Drech}. These difficulties are compounded with the secondary star not being spectroscopically visible. Also the formation of circumbinary objects remains an open question. If they do exist, were they formed before or after the common envelope ejection, or a combination of both? (See for example  \cite{Schlei}).\\

\noindent 
These uncertainties can make it difficult to interpret observed eclipse time variations, particularly when time lines are short.  Our recent observations of seven sdB binary systems, together with historical data, indicate that:\\
\begin{itemize}
\renewcommand{\labelitemi}{\scriptsize$\bullet$}
\item 
From earlier eclipse time observations, it has been suggested that five  (HS0705+6700, HS2231+2441, NSVS 07826147, NSVS 14256825 and NY Vir) of the seven systems reported herein most probably had circumbinary companions. More recent observations indicate only three of these systems, HS0705+6700, NSVS 14256825 and NY Vir, show significant LTT effects.  All three have recently shown very significant changes from their earlier circumbinary hypothesis. HS0705+6700, after more than 1.5 circumbinary periods, now gives little indication of eclipse time variations. It is evident that observation of several consecutive and consistent LTT cycles is a prerequisite to support a third body hypothesis.\\
\noindent
\item Whilst NY Vir and NSVS 14256825 have shown significant changes from earlier eclipse time predictions both still show a cyclical shape to their (O – C) residuals. In particular the residuals of NSVS 14256825 give a very close fit to the more recent predicted parameters of \cite{Nasi}.  However, both systems report only one cycle of a proposed circumbinary companion and drawing firm conclusions would thus be premature.
\item
Four systems, EC10246-2707, HS2231+2441, J0802+0008 and NSVS 07826147, show little, or no, eclipse time variations. Linear and quadratic ephemerides have been determined for each system and although all four systems tended to give marginally lower residuals with the quadratic ephemeris, statistically there was no significant difference between either ephemeris.\\
\item\noindent 
We find agreement with \cite{bours2016long} that a decade or so of observations are required to establish a reliable ephemeris but we do not find a tendency for the RMS weighted residuals to saturate at $\sim$100. For the systems we considered we found RMS values ranging between 1.6 and 21.0. We also found only one of the seven systems, HS0705+6700, had (O – C) residuals approaching 100s with the other systems falling significantly short of this target. We found that the RMS value was strongly dependent upon the LTT amplitude and inversely dependent upon the uncertainty in minima. Systems with a few abnormally low uncertainty values can significantly bias upwards the magnitude of the RMS weighted residuals.\\
\noindent

\item
Although the spectral types of sdB companions are not tightly constrained, we are, however, in general agreement with Bours that larger values of RMS of residuals are generally found with companions of spectral type M5/6 or earlier. This is the boundary where a star becomes fully convective and may be indicative that significant LTT effects are being driven by magnetic processes within the companion star, possibly from a modified Applegate type mechanism.  There are some possible anomalies, e.g. EC10246-2707 and NSVS 07826147, which are both classified as M5 but show little LTT effects.\\

\end{itemize}
A longer time baseline is required to resolve underlying nature of these eclipse time variations, but so far only NY Vir has been included in the NASA Exoplanet archive.  Similarly for short period white dwarf eclipsing binaries only NN Ser, DP Leo, RR Cae and UZ For have confirmed exoplanets.

\begin{acknowledgements}
This work was part funded by the British Astronomical Association under BAA account BAAHS0705.  This work also makes use of observations from the LCO network of telescopes and of the APASS database maintained on the AAVSO website .  We would like to thank Dr. Marcus Lohr (The Open University) who provided information on the methodology used in analysing SuperWASP data and Dr. David Boyd for providing two sets of observations that enabled us to link our new data with the historical observations of NY Vir.  We would also like to thank Dr. Brad Barlow (High Point University), Dr. Klaus Beuermann  (University of Goettingen), Dr. Roy Østensen (University of Leuven), Dr. Stephan Geier (University of Tübingen), Dr Veronika Schaffenroth (University of Innsbruck), Patrick Wils (Vereniging voor Sterrenkunde), Dr Slowikowska (University of Zielona Góra) and Dr. S-B Qian (Chinese Academy of Sciences) who addressed many of the questions we posed. We would also like to thank the referee whose guidance  was greatly appreciated. 
\end{acknowledgements}

\begingroup
\setlength{\bibsep}{0pt plus 0.3ex}
     \bibliographystyle{aa} 
    \bibliography{aaref2} 
\endgroup


\cleardoublepage
\clearpage


\appendix
\Large
\noindent\textbf{Appendix}\\\\
\normalsize
\noindent\textbf{Note} The full tables are available on-line.
\begin{table}[H]
\small
\renewcommand\thetable{A.1}
\caption{Telescopes and instrumentation used for the measurements reported in this paper. 
}             
\label{table:A.1}      
\begin{tabular}{llll}        

\hline \\   

Observatory & Telescope & Instrumentation &  MPC Code \\
\\
\hline\\
Sierra Stars Observatory Network&0.61m F/10 Optical&Finger Lakes Inst& G68\\
Markleeville&Mechanics&ProLine camera&\\
California, USA &Nighthawk CC06&3056 x 3056 pixels&\\
http://sierrastars.com/gp/SSO/SSO-CA.aspx& &FOV 21 x 21 arcmin&\\
\hline
\end{tabular}
\end{table}

\begin{table}[H]
\small
\renewcommand\thetable{A.2}
\caption{Compilation of our new measurements observed between 2013 September and 2017 July.  The reference epoch is noted for each binary system in each section header.
}             
\label{table:A.2}      
\begin{tabular}{lcccll}       
\hline
\\
BJD & Error & Cycle & Minima&Filter&Telescope\\ 
& (days)& & & & \\\hline
EC 10246-2707 &  $T_{0}$= 2450493.46733 &  &  &  & \\
 \hline
2457407.697910 & 0.000019 & 58344 & I & V Bessell & 1m Cerro Tololo, Chile, LCO  \\ 
	2457408.645870 & - & 58352 & I & B Bessell & 1m Cerro Tololo, Chile, LCO \\ 
	2457425.592548 & 0.000015 & 58495 & I & V Bessell & 1m Sutherland, SA, LCO  \\ 
	2457436.139746 & 0.000088 & 58584 & I & Sloan r' & 0.43m iTelescope, Siding Spring T17 \\ 
	2457449.531205 & 0.000022 & 58697 & I & Unfiltered & 1m Sutherland, SA, LCO  \\ 
	2457775.072622 & 0.000091 & 61444 & I & Sloan r' & 0.43m iTelescope, Siding Spring T17\\ 
	2457777.087370 & 0.000084 & 61461 & I & Sloan r' & 0.43m iTelescope, Siding Spring T17 \\ 
    \hline
HS0705+6700 & $T_{0}$ = 2451822.76155 &  &  &  &  \\ 
\hline
\end{tabular}
\end{table}

\begin{table}[H]
\small
\renewcommand\thetable{A.3}
\caption{Comparison stars coordinates and apparent magnitudes from APASS catalogue. The star reference numbers relates back to star charts used for this work. 
}             
\label{table:A.3}      
\begin{tabular}{lclllllllc}       
\hline  
\\
Binary & Star & Johnson V & Johnson B&Sloan g'&Sloan r'& Sloan i'&RA&Dec&Distance from\\ 

System& Ref No& & & & &&&&Target (arcmin) \\
\\ \hline
\\
EC10246-2707& 1 & 13.273 & 13.699 & 13.443 & 13.158 & 13.079 & 10:27:02.841 & -27:21:02.57 & 2.5 \\ 

RA   10:26:56.472& 2 & 14.220 & 14.750 & 14.427 & 14.052 & 13.943 & 10:26:57.964 & -27:27:27.17 & 4.51 \\ 
Dec -27:22:57.11 	 & 4 & 13.460 & 13.898 & 13.604 & 13.271 & 13.179 & 10:27:14.648 & -27:23:28.54 & 4.58 \\ 
	 & 5 & 14.701 & 15.253 & 14.914 & 14.521 & 14.352 & 10:27:07.271 & -27:18:48.15 & 4.96 \\ 
	 & 6 & 15.225 & 15.795 & 15.506 & 15.153 & 15.038 & 10:27:13.378 & -27:16:18.34 & 7.88 \\ 
	 & 7 & 14.500 & 15.166 & 14.779 & 14.243 & 14.118 & 10:26:24.711 & -27:18:01.56 & 9.34 \\ \hline
\\
HS0705+6700&1 & 13.637 & 14.141 & 13.824 & 13.514 & 13.358 & 07:11:21.473 &   67:00:57.55 & 11.18 \\
RA   07:10:42.056	 & 2 & 13.435 & 14.172 & 13.728 & 13.190 & 12.967 & 07:09:10.503 &   66:58:51.25 & 23.08 \\ 
Dec 66:55:43.52	 & 3 & 13.798 & 14.230 & 13.947 & 13.685 & 13.574 & 07:09:24.091 &   67:01:45.42 & 20.38 \\ 
	 & 4 & 13.842 & 14.284 & 14.001 & 13.757 & 13.636 & 07:10:24.685 &   66:56:18.49 & 4.36 \\ 
	 & 6 & 14.135 & 14.488 & 14.239 & 14.047 & 13.946 & 07:09:47.389 &   66:55:16.33  & 13.65 \\ 
\end{tabular}
\end{table}
\clearpage


\end{document}